\documentclass[12pt]{article}
\usepackage{epsfig,amssymb,amsmath,psfrag}

\catcode`\@=11
\def \be  {\begin{equation}}
\def \ee  {\end{equation}}
\def \ba  {\begin{eqnarray}}
\def \ea  {\end{eqnarray}}
\def \baa {\begin{eqnarray*}}
\def \eaa {\end{eqnarray*}}
\def \bb  {\begin {thebibliography} }
\def \eb  {\end{thebibliography}}
\def \lab #1 {\label{#1}}
\newcommand{\ft}[2]{{\textstyle\frac{#1}{#2}}}

\newcommand\re[1]{(\ref{#1})}

\def \qqqquad {\qquad\qquad}
\def \matrix #1 {\left(\begin{array}{cc} #1 \end{array}\right)}

\def \tr {\mathop{\rm tr}\nolimits}

\def \e  {\mathop{\rm e}\nolimits}

\newcommand\lr[1]{{\left({#1}\right)}}

\newcommand \vev [1] {\langle{#1}\rangle}

\newcommand{\as}{\ifmmode\alpha_{\rm s}\else{$\alpha_{\rm s}$}\fi}
\newcommand{\asbar}{\ifmmode\bar{\alpha}_{\rm s}\else{$\bar{\alpha}_{\rm s}$}\fi}

\font\cmss=cmss12 
\def\inbar{\,\vrule height1.5ex width.4pt depth0pt}
\def\IC{\relax\hbox{$\inbar\kern-.3em{\rm C}$}}
\def\IZ{\relax{\hbox{\cmss Z\kern-.4em Z}}}
\def\IR{{\hbox{{\rm I}\kern-.2em\hbox{\rm R}}}}

\def\IP{{\hbox{{\rm I}\kern-.2em\hbox{\rm P}}}}
\def\II{\hbox{{1}\kern-.25em\hbox{l}}}

\def \vep {\epsilon}

\begin{document}

%%%%%%%%%
% Front page here
\thispagestyle{empty}
%\vspace*{1cm}
\null\vskip-12pt \hfill  LAPTH--1240/08 \\
\null\vskip-12pt \hfill LPT--Orsay--08--29
%\vskip-10pt \hfill {\tt hep-th/??????}
\vskip2.2truecm
\begin{center}
\vskip 0.2truecm {\Large\bf
%\titleline
{\Large Hexagon Wilson loop = six-gluon MHV amplitude}\\
} \vskip 1truecm
%\vfill
{\bf J.M. Drummond$^{*}$, J. Henn$^{*}$, G.P. Korchemsky$^{**}$ and E. Sokatchev$^{*}$ \\
}

\vskip 0.4truecm
%\addresses
$^{*}$ {\it LAPTH \footnote{Laboratoire de Physique Th\'{e}orique d'Annecy-le-Vieux, UMR 5108}, Universit\'e de Savoie, CNRS; \\ 9 Chemin de Bellevue, B.P. 110, F-74941 Annecy-le-Vieux Cedex, France\\
\vskip .2truecm $^{**}$ {\it
Laboratoire de Physique Th\'eorique%
\footnote{Unit\'e Mixte de Recherche du CNRS (UMR 8627)},
Universit\'e de Paris XI, \\
F-91405 Orsay Cedex, France
                       }
  } \\

\end{center}

\vskip 1truecm %\Large
%\noindent
\centerline{\bf Abstract} % \normalsize

\medskip

\noindent

%\centerline{\today}
We compare the two-loop corrections to the finite part of the light-like hexagon Wilson loop with
the recent numerical results for the finite part of the MHV six-gluon amplitude in ${\mathcal N}=4$
SYM theory by Bern, Dixon, Kosower, Roiban, Spradlin, Vergu and Volovich (arXiv:0803.1465 [hep-th])
and demonstrate that they coincide within the error bars and, at the same time, they differ from
the BDS ansatz by a non-trivial function of (dual) conformal kinematical invariants. This provides
strong evidence that the Wilson loop/scattering amplitude duality holds in planar ${\mathcal N}=4$
SYM theory to all loops for an arbitrary number of external particles.

\newpage
\setcounter{page}{1}\setcounter{footnote}{0}

\section{Introduction}

In an important recent development in the study of the AdS/CFT correspondence, Alday and Maldacena
proposed \cite{am07} the strong coupling description of planar gluon scattering amplitudes in the
maximally supersymmetric $\mathcal{N}=4$ Yang-Mills (SYM) theory and were able to make a direct
comparison with a prediction based on weak coupling results for the same amplitudes. At weak
coupling, following years of intensive studies of gluon scattering amplitudes
\cite{Bern:1994zx,Bern:1997nh,Bern:2007dw}, the conjecture was  put forward by Bern, Dixon and
Smirnov \cite{bds05} (see also an earlier version in \cite{Anastasiou:2003kj}) that the maximally
helicity-violating (MHV) planar gluon amplitudes in $\mathcal{N}=4$ SYM have a remarkably simple
all-loop iterative structure. In general, these amplitudes have the following form:
\begin{equation}\label{BDS'}
\ln\mathcal{M}_n^{\rm
  (MHV)} = \text{[IR divergences]} + F_n^{\rm
  (MHV)}(p_1,\ldots,p_n; a) + O(\epsilon)\,.
\end{equation}
{Here $\mathcal{M}_n^{\rm (MHV)}$ is the color-ordered planar gluon amplitude, divided by the tree
amplitude.} The first term on the right-hand side describes the infrared (IR) divergences and the
second term is the finite contribution dependent on the gluon momenta $p_i$ and on the 't Hooft
coupling $a= {g^2 N}/(8\pi^2)$. The structure of IR divergences is well understood in any gauge
theory~\cite{IR}. In particular, in theories with a vanishing beta function like ${\cal N}=4$ SYM,
the leading IR singularity in dimensional regularization is a double pole, whose coefficient is the
universal cusp anomalous dimension appearing in many physical processes
\cite{KR87,KM93,Korchemskaya:1996je,Belitsky:2003ys}. The BDS conjecture provides an explicit
expression for the finite part, $F_n^{\rm (MHV)}=F_n^{\rm (BDS)}$, for an arbitrary number $n$ of
external gluons, to all orders in the coupling $a$. Remarkably, the dependence of $F_n^{\rm (BDS)}$
on the kinematical invariants is described by a function which is coupling independent and,
therefore, can be determined at one loop. At present, the BDS conjecture has been tested up to
three loops for $n=4$ \cite{bds05} and up to two loops for $n=5$ \cite{5point}. The explicit
investigation of the conjecture for $n=6$ at two loops is the subject of this paper and of the
parallel paper \cite{parallel}.

As mentioned earlier, after the work of Alday and Maldacena it became possible to test the BDS
conjecture at strong coupling. According to their proposal, at strong coupling the planar gluon
amplitude is related to the area of a minimal surface in AdS${}_5$ space attached to a specific
closed contour $C_n$, made out of $n$ light-like segments $[x_i,x_{i+1}]$ defined by the gluon
momenta $x_i^\mu-x_{i+1}^\mu = p_i^\mu$ (with the cyclicly condition $x_{n+1} \equiv x_1$),
\begin{equation}\label{AM}
\ln \mathcal{M}_n = -\frac{\sqrt{g^2 N}}{2\pi} A_{\rm min}(C_n)\,.
\end{equation}
For $n=4$ the minimal surface $A_{\rm min}(C_4)$ was found explicitly in \cite{am07}, by making use
of the conformal symmetry of the problem. With the appropriate AdS equivalent of dimensional
regularization, the divergent part of $\ln \mathcal{M}_4$ has the expected pole structure, with the
coefficient in front of the double pole given by the known strong coupling value of the cusp
anomalous dimension. Most importantly, the finite part of $\ln \mathcal{M}_4$ is in perfect
agreement with $F_4^{\rm (BDS)}$ from the BDS ansatz. For $n\ge 5$ the practical evaluation of the
solution of the classical string equations turns out to be difficult,  but it simplifies
significantly for $n$ large \cite{Alday:2007he}. In the limit $n\to\infty$ the strong coupling
prediction for $\ln\mathcal{M}_n$ disagrees with the BDS ansatz. This indicates \cite{Alday:2007he}
that the BDS conjecture should fail for a sufficiently large number of gluons and/or at
sufficiently high loop level.

Alday and Maldacena pointed out \cite{am07} that their prescription \re{AM} is mathematically
equivalent to the strong coupling calculation of the expectation value of a Wilson loop $W(C_n)$,
defined on the light-like contour $C_n$ \cite{M98,Kr02}. This should not come as a total surprise,
since the intimate relationship between the infrared divergences of the scattering of massless
particles and the ultraviolet divergences of Wilson loops with cusps is well known in QCD
\cite{KR87,KM93,Korchemskaya:1996je}. Inspired by this, in \cite{DKS07} three of us conjectured
that a similar duality relation between planar gluon amplitudes and light-like Wilson loops also
exists at weak coupling. We illustrated this duality by an explicit one-loop calculation in the
simplest case $n=4$. This was later extended to the case of arbitrary $n$ at one loop in
\cite{BHT07}.  If we write the log of the Wilson loop in the following way,
\begin{equation}\label{logWL}
\ln W(C) = \text{[UV divergences]} + F_n^{\rm
  (WL)}(x_1,\ldots,x_n; a) + O(\epsilon)\,,
\end{equation}
then the duality relation identifies the finite parts of the two objects up to an additive constant, once one imposes the relations $p_i^\mu = x_i^\mu - x_{i+1}^\mu$,
\begin{equation}\label{finiteduality}
F_n^{\rm (MHV)} = F_n^{\rm (WL)} +\text{const}\ .
\end{equation}
This property is extremely non-trivial.

Further evidence in favor of the duality relation \re{finiteduality} at weak coupling came from
two-loop calculations of $\ln W(C_n)$ for $n=4$ \cite{Drummond:2007cf} and $n=5$
\cite{Drummond:2007au}. Our results were in agreement with the two-loop MHV gluon amplitude
calculations \cite{Anastasiou:2003kj,5point}, and hence with the BDS ansatz for $n=4,5$.
Furthermore, in \cite{Drummond:2007cf} we proposed and in \cite{Drummond:2007au} we proved a
conformal Ward identity for the light-like Wilson loop $W(C_n)$, valid to all orders in the
coupling. It fixes the functional form of the finite part of $\ln W(C_n)$ for $n=4$ and $n=5$, up
to an additive constant, to agree with the conjectured BDS form for the corresponding gluon amplitudes.%
\footnote{Later, similar Ward identities were also obtained at strong coupling using the AdS/CFT
correspondence in Refs.~\cite{AM-un,Komargodski:2008wa}.} However, for $n\ge 6$, although the BDS ansatz $F_n^{\rm (BDS)}$ satisfies the conformal Ward identity, $F_n^{\rm (WL)}$ is allowed to differ from $F_n^{\rm (BDS)}$ by an
arbitrary function of conformal invariants (for $n=6$ there are three such invariants). This result
provided a possible explanation of the BDS conjecture for $n=4,5$ (assuming that the MHV amplitudes
have the same conformal properties as the Wilson loop, see the discussion in Sect.~2.4 below), but
left the door open for potential deviations from it for $n\geq 6$. To verify whether the BDS
conjecture and/or the proposed duality relation \re{finiteduality} still hold for $n=6$ to two
loops, it was necessary to perform explicit two-loop calculations of the finite parts of the
six-gluon amplitude $F_6^{\rm (MHV)}$, and of the hexagon Wilson loop $F_6^{\rm (WL)}$.

In a recent paper \cite{Drummond:2007bm} we reported on the two-loop calculation of $F_6^{\rm
(WL)}$ and found that it differs from the BDS ansatz,
\begin{equation}\label{discrep}
  F_6^{\rm (WL)} = F_6^{\rm (BDS)}+R_6\,.
\end{equation}
Here, in complete agreement with the conformal Ward identity, $R_6$ is a non-trivial  `remainder'
function of three conformally invariant combinations of the kinematical variables. {As was
emphasized in \cite{Drummond:2007bm}, were the duality relation \re{finiteduality} to hold for $n=6$, the
function $R_6$ would describe the discrepancy between the BDS ansatz and the scattering amplitude.}
A parallel two-loop six-gluon amplitude calculation has been undertaken by Bern, Dixon, Kosower,
Roiban, Spradlin, Vergu and Volovich and the results became available very recently
\cite{parallel}.

The detailed numerical comparisons of the two calculations, described in this paper and in the
parallel publication \cite{parallel}, shows that, firstly, the BDS ansatz fails for $n=6$ at two
loops and, secondly, the {\it duality with Wilson loops is preserved},
\begin{equation} %\label{duality-7}
F_6^{\rm (MHV)}=F_6^{\rm (WL)}- c_6(a)\,,
\end{equation}
{where $c_6(a)$ is a constant.} We consider this as very strong evidence that the duality relation
\re{finiteduality} should hold for arbitrary $n$ to all orders in the coupling.

The real challenge now is, in our opinion, to find out the deep reason behind the surprising
duality \re{finiteduality} between two apparently unrelated objects in the ${\cal N}=4$ theory. One might
speculate that the scattering amplitudes and the light-like Wilson loops share the same (probably
infinite) set of symmetries, of which (dual) conformal symmetry is just the most visible part. If
so, these symmetries may completely fix the form of the quantities on both sides of the duality
relation. This could be the manifestation of some new type of integrability of $\mathcal{N}=4$ SYM
theory. We would like to stress that the duality relation \re{finiteduality} only holds in the planar
limit. Indeed,  the known two-loop non-planar contributions to the four-gluon amplitude
\cite{Bern:1997nh} appear to break dual conformal symmetry, which, as we have shown, is an
important ingredient of the duality. Also, recent studies in the AdS/CFT correspondence
suggest~\cite{McGreevy:2007kt,Komargodski:2007er,Oz:2007qr} that at strong coupling the duality
relation \re{finiteduality} can be extended to scattering amplitudes involving gluinos (both in the
$\mathcal{N}=4$ SYM and its deformed versions), as well as to the on-shell matrix elements of
conserved currents~\cite{Alday:2007he}. It would be interesting to verify whether the same relation
holds at weak coupling. This question is beyond the scope of the present paper and deserves further
investigation.

We would like to point out that a weaker form of the duality \re{finiteduality} has already been observed
in QCD in the special, high-energy (Regge) limit $s\gg -t >0$ for the four-gluon amplitude up to
two loops ~\cite{Korchemskaya:1996je}. The same relationship holds in any gauge theory ranging from
QCD to $\mathcal{N}=4$ SYM. The essential difference between these theories is that in the former
case the duality is only valid in the Regge limit, whereas in the latter case it is exact in
general kinematics. Moreover, in Ref.~\cite{DKS07}, based on the BDS ansatz, three of us argued
that the four-gluon amplitude is Regge exact in $\mathcal{N}=4$ SYM. Namely, the contribution of
the gluon Regge trajectory to $\ln {\cal M}_4$ coincides with its exact expression evaluated for
arbitrary values of $s$ and $t$, up to terms vanishing as $\epsilon\to 0$. This property allowed us
to obtain the explicit expression for the three-loop gluon Regge trajectory in $\mathcal{N}=4$
SYM~\cite{DKS07}.  The two-loop correction to this trajectory was found to be in agreement with the
results of Ref.~\cite{Kotikov:2000pm} and the three-loop correction was later confirmed in
Refs.~\cite{Naculich:2007ub,Brower:2008nm,Bartels:2008ce,DelDuca:2008pj}. We would like to mention
that a thorough analysis of the Regge limit of planar multi-gluon amplitudes in $\mathcal{N}=4$ SYM
was recently performed in Refs.~\cite{Brower:2008nm,Bartels:2008ce} and it provided further
evidence that the BDS ansatz needs to be corrected~\cite{Bartels:2008ce}.

The paper is organized as follows. In Section \ref{pgawld} we explain in detail the proposed
duality between planar gluon amplitudes and light-like Wilson loops.  We summarize some basic facts
about planar gluon amplitudes, focusing on the structure of their infrared divergences and on the
BDS ansatz for the finite part. Next we describe the light-like Wilson loop, the structure of its
ultraviolet singularities and the anomalous conformal Ward identity for its finite part. Then we
state the duality relation and discuss its consequences for the six-gluon amplitude. Section 3
contains a description of our two-loop calculation of the hexagon Wilson loop $W(C_6)$ (already
reported in \cite{Drummond:2007bm}). We separate the divergent and finite part of individual
Feynman diagrams by employing the `subtraction procedure' proposed in \cite{Drummond:2007au}, and
then demonstrate that the divergent part of $\ln W(C_6)$ is of the expected form. We work out the representation for the finite part
$F_6^{\rm (WL)}$ in the form of convergent multiple parameter integrals which can easily be
evaluated numerically for given kinematical configurations. In Section \ref{drfsgma} we perform a
detailed numerical comparison of our results of the hexagon Wilson loop calculation with the BDS
ansatz for $n=6$ and with the results of the parallel six-gluon calculation \cite{parallel}.
Section~5 contains concluding remarks.

\section{Planar gluon amplitude/Wilson loop duality}\label{pgawld}

Recent studies revealed that the gluon scattering amplitudes have a number of remarkable properties
in the $\mathcal{N}=4$ SYM theory. To describe them, we first recall some general features of such
amplitudes.

\subsection{Planar amplitudes}\label{pa}

In a generic Yang-Mills theory with an $SU(N)$ gauge group, the scattering amplitude of $n$ gluons
can be decomposed into color-ordered partial amplitudes multiplied by the corresponding color
structure~\cite{Mangano:1990by}. In the planar limit, the dominant contribution only comes from the
single-trace color structures leading to
\begin{align}\label{A-planar}
\mathcal{A}_n(\{p_i,h_i,a_i\}) = 2^{n/2} g^{n-2} \sum_{\sigma\in S_n/Z_n} \tr
[t^{a_{\sigma(1)}}\ldots t^{a_{\sigma(n)}} ]  A_n\left(\sigma(1^{h_1},\ldots,
n^{h_n})\right) + O(1/N^2),
\end{align}
where each gluon is characterized by its on-shell momentum $p_i^\mu$ ($p_i^2=0$), helicity $h_i=\pm
1$ and color index $a_i$. Here the sum runs over all possible non-cyclic permutations $\sigma$ of
the set $\{1,\ldots,n\}$ and the color trace involves the generators $t^a$ of $SU(N)$ in the
fundamental representation normalized as $\tr(t^a t^b) =\ft12\delta^{ab}$. All gluons are treated
as incoming, so that the momentum conservation takes the form $\sum_{i=1}^n p_i=0$.

\begin{figure}
\psfrag{=}[cc][cc]{$\Longleftrightarrow$}\psfrag{dots}[cc][cc]{$\bf\ldots$}
\psfrag{x1}[cc][cc]{$x_1$} \psfrag{x2}[cc][cc]{$x_2$} \psfrag{x3}[cc][cc]{$x_3$}
\psfrag{x4}[cc][tc]{$\ddots $} \psfrag{x5}[cc][cc]{$x_{n-1}$} \psfrag{x6}[rc][cc]{$x_n$}
\psfrag{p1}[cc][cc]{$p_1$} \psfrag{p2}[cc][cc]{$p_2$} \psfrag{p3}[cc][cc]{$p_3$}
\psfrag{p4}[cc][bc]{$\vdots $} \psfrag{p5}[cc][cc]{$p_{n-1}$} \psfrag{p6}[cc][cc]{$p_n$}
\centerline{{\epsfysize5.5cm \epsfbox{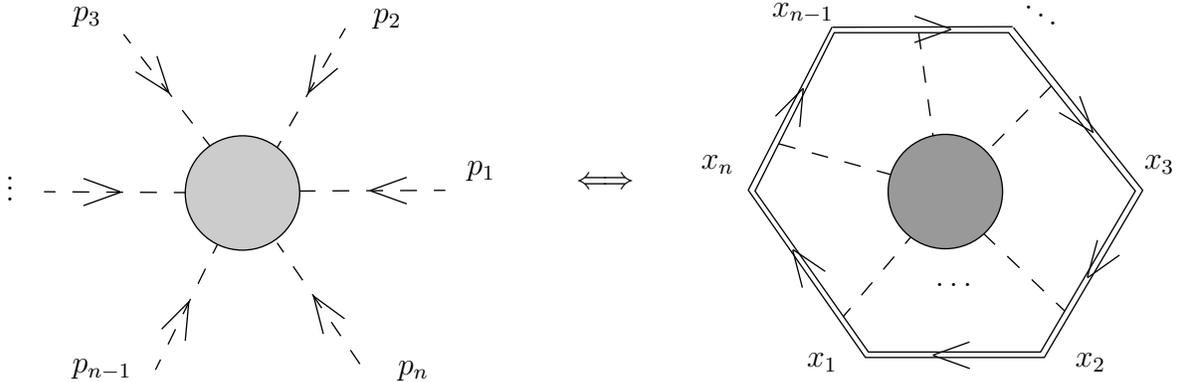}}} \caption[]{\small The conjectured duality
relation between the gluon scattering amplitude $\mathcal{M}_n$ and the Wilson loop  $W(C_n)$. The
dashed lines depict gluons and the double lines the integration contour $C_6$. The momenta of the
incoming gluons are identified as the light-like segments of the integration contour, $p_i \sim x_i
- x_{i+1}$. } \label{Fig:example}
\end{figure}

The planar scattering amplitude \re{A-planar} is uniquely determined by the set of color-ordered
amplitudes $A_n$. It follows from the supersymmetric Ward identities \cite{Grisaru:1976vm} that the
amplitude $A_n$ vanishes to all orders when either all the external gluons or all gluons but one
have the same helicity. When two gluons have, e.g., negative helicity and the remaining $n-2$
gluons have positive helicity, the so-called maximally helicity-violating (MHV) amplitudes have the
following remarkably simple form in the spinor helicity formalism:
\begin{equation}\label{MHV}
A_n(1^+\ldots m_1^-\ldots m_2^-\ldots n^+) = i \frac{\vev{m_1\,
m_2}^4}{\vev{1\, 2}\vev{2\, 3}\ldots \vev{n\, 1}} \mathcal{M}_n^{\rm (MHV)}\,.
\end{equation}
Here the helicity structure is described by  massless Weyl spinors $u_\pm(p)$ of momentum $p$ and
chirality $\pm 1$; they form Lorentz invariant `inner products' of the type $\vev{j\, k} =
\vev{j^-| k^+} = \bar u_-(p_j) u_+(p_k)$. The Lorentz scalar $\mathcal{M}_n = 1 + a
\mathcal{M}_{n;1} + O(a^2)$ does not depend on the positions $m_{1,2}$ of the negative-helicity
gluons~\cite{BDDK}. According to \re{A-planar} and \re{MHV}, the perturbative corrections to the
planar MHV amplitude are determined to \textit{all loops} by the single function
$\mathcal{M}_n^{\rm (MHV)}$ of the Mandelstam variables. This function (more precisely, its
logarithm) is the main object of interest in this paper.

It is worthwhile mentioning that for scattering amplitudes with more than two negative helicities
(next-to-MHV amplitudes and their generalizations) the situation is more complex. At tree level, in
the spinor helicity formalism these amplitudes are expressed in terms of various Lorentz structures
built from spinors~\cite{Mangano:1990by}. At one-loop level, however, the perturbative corrections
to the non-MHV amplitudes induce new Lorentz structures~\cite{Bern:2007dw}.  As a result, in
distinction with \re{MHV}, the non-MHV amplitudes do not admit a simple factorized form. This is
the reason why the duality \re{finiteduality} with Wilson loops that we discuss here only concerns MHV
amplitudes.

\subsection{Planar MHV amplitudes}\label{pmhva}

The gluon scattering amplitudes suffer from infrared divergences. In supersymmetric gauge theories,
they can be regularized using the dimensional reduction scheme (DRED) with $D=4-2\epsilon_{\rm IR}$
and $\epsilon_{\rm IR}<0$. \footnote{More precisely, a modification of the DRED scheme, the
so-called four-dimensional helicity (FDH) scheme, has been used in~\cite{FDH}.} In this case, the
IR divergences appear as poles in $\epsilon_{\rm IR}$.  At a given loop order $l$, the maximal
order of the poles is $2l$. In a generic Yang-Mills theory, the IR divergent part of the planar
scattering amplitudes has a universal form which is sensitive neither to the helicities of the
scattered particles, nor to their type (gluons, fermions, scalars). More precisely, the IR
divergences exponentiate in the all-loop planar amplitude and, as a consequence, they can be
factorized into a universal divergent factor. In application to the MHV planar amplitudes in
$\mathcal{N}=4$ SYM theory, this property allows one to decompose $\ln \mathcal{M}_n^{\rm (MHV)}$
into  divergent and finite parts as follows,
\begin{equation}\label{M=Z+F}
 \ln \mathcal{M}_n^{\rm (MHV)} = Z_n + F_n^{\rm (MHV)} + O(\epsilon_{\rm \scriptscriptstyle IR})\,,
\end{equation}
In a theory with a vanishing beta function, the IR divergent part $Z_n$ takes the particularly simple form
\begin{equation}\label{Zn}
Z_n = -\frac14 \sum_{l\geq 1} a^l \lr{\frac{\Gamma^{(l)}_{\rm cusp}}{(l\epsilon_{\rm
\scriptscriptstyle IR})^2}+\frac{G^{(l)}}{l\epsilon_{\rm \scriptscriptstyle IR}}}\sum_{i=1}^n
\lr{-\frac{t_i^{[2]}}{\mu_{\rm\scriptscriptstyle IR}^2}}^{-l\epsilon_{\rm \scriptscriptstyle IR}}\,,
\end{equation}
where $t_i^{[2]} \equiv s_{i,i+1} = (p_i+p_{i+1})^2$ is the invariant mass of two adjacent gluons
with indices $i$ and $i+1$ subject to the periodicity condition $i+n\equiv i$. The IR cut-off
$\mu_{\rm\scriptscriptstyle IR}^2$ is related to the dimensional regularization scale $\mu^2$ as
\begin{equation}\label{ge}
\mu_{\rm\scriptscriptstyle IR}^2 =4\pi{\rm e}^{-\gamma_E}\mu^2\,,
\end{equation}
where $\gamma_E$ is the Euler constant.

We did not put the superscript MHV on $Z_n$ in order to indicate that it has a universal form for
all planar amplitudes in the $\mathcal{N}=4$ SYM theory. The coefficients of the IR poles in
\re{Zn} are the expansion coefficients of the cusp anomalous dimension $\Gamma_{\rm cusp}(a) =
\sum_{l=1}^\infty a^l \Gamma^{(l)}_{\rm cusp}$ and of the so-called collinear anomalous dimension
$G(a)= \sum_{l=1}^\infty a^l G^{(l)}$. To two-loop order, they are given in the DRED scheme by
\begin{align}\label{cusp}
\Gamma_{\rm cusp}(a) = 2a -2\zeta_2 a^2 +O(a^3)
\,,\qquad
G(a) = -\zeta_3 a^2 + O(a^3)\,.
\end{align}
By definition~\cite{P80,KR87,KR86}, the cusp anomalous dimension $\Gamma_{\rm cusp}(a)$ describes
specific \textit{ultraviolet} divergences of a Wilson loop evaluated over a contour with a cusp.
Its appearance in the \textit{infrared} divergent part of the scattering amplitude \re{Zn} is not
accidental, of course. It has its roots in the deep relation between scattering amplitudes in gauge
theory and Wilson loops evaluated over specific contours in Minkowski space-time, defined by the
particle momenta \cite{KR87,KM93,Korchemskaya:1996je}. It should be mentioned that this relation is
not specific to $\mathcal{N}=4$ SYM and it holds in any gauge theory, including QCD. The two-loop
expression for $\Gamma_{\rm cusp}(a)$ in a generic (supersymmetric) Yang-Mills theory was found in
Refs.~\cite{KR86,Belitsky:2003ys}. We remark that in $\mathcal{N}=4$ SYM theory, $\Gamma_{\rm
cusp}(a)$ is known at weak coupling to four loops \cite{Bern:2006ew}, and there is a conjecture for
it to all loops \cite{Beisert:2006ez}. At strong coupling, the solution to the BES equation
proposed in \cite{Beisert:2006ez} produces a strong coupling expansion of $\Gamma_{\rm
cusp}(a)$~\cite{Benna:2006nd,Basso:2007wd,Kostov:2008ax}. The first few terms of this expansion are
in agreement with the existing quantum superstring calculation of
Refs.~\cite{Gubser:2002tv,Frolov:2002av,Roiban:2007dq}. The non-universal collinear anomalous
dimension $G(a)$ is known to four loops at weak coupling \cite{bds05,Cachazo:2007ad}.

Let us now examine the finite part of the MHV planar amplitude $F_n^{\rm (MHV)}$.  In a generic
Yang-Mills theory, it has a much more complicated form compared to the divergent part.
Surprisingly, this finite part becomes considerably simpler in ${\cal N}=4$ SYM. Previously, the
finite parts of the following MHV planar amplitudes have been calculated:
\begin{itemize}
  \item $n=4$ gluon amplitude up to three loops \cite{Bern:1997nh,Anastasiou:2003kj,bds05};
  \item $n=5$ gluon amplitude up to two loops \cite{5point};
  \item $n\ge 6$ gluon amplitude at one loop \cite{Bern:1994zx}.
\end{itemize}
These calculations revealed a remarkable iterative structure of $F_n^{\rm (MHV)}$. They led to the
formulation of the Bern-Dixon-Smirnov ansatz (BDS) which provides a conjectured expression for the finite part
of the MHV planar amplitudes valid for an arbitrary number of gluons $n\ge 4$ and to all orders in
the coupling. The BDS ansatz reads
\begin{equation}\label{BDS}
F_n^{\rm (BDS)} = \frac12\Gamma_{\rm cusp}(a) \mathcal{F}_n + c(a),
\end{equation}
where $\Gamma_{\rm cusp}(a)$ is the cusp anomalous dimension defined in \re{cusp} and $c(a)
=-\frac12 \zeta_2^2 a^2 + O(a^3)$ is a universal (independent of $n$) additive constant. The
dependence on the kinematical invariants is described by the function $\mathcal{F}_n$.  According
to the BDS conjecture, this function is coupling independent, and can thus be determined at one
loop. For our purposes here we only need its explicit expressions for $n=4,5,6$:
\begin{align}\label{f4}
\mathcal{F}_4 &= \frac{1}{2} \ln^2 \Bigl( \frac{t_1^{[2]}}{t_2^{[2]}} \Bigr) + 4 \zeta_2,
\\\label{f5}
\mathcal{F}_5 &= \frac{1}{2} \sum_{i=1}^5 \biggl[ -\frac{1}{2} \ln \Bigl(
\frac{t_i^{[2]}}{t_{i+3}^{[2]}} \Bigr) \ln \Bigl( \frac{t_{i+1}^{[2]}}{t_{i+2}^{[2]}}\Bigr) +
\frac{3}{2} \zeta_2 \biggr],
\\ \label{f6}
\mathcal{F}_6 &= \frac{1}{2} \sum_{i=1}^6 \biggl[ -\ln \Bigl( \frac{t_i^{[2]}}{t_i^{[3]}} \Bigr)
\ln \Bigl( \frac{t_{i+1}^{[2]}}{t_i^{[3]}} \Bigr) + \frac{1}{4} \ln^2 \Bigl(
\frac{t_i^{[3]}}{t_{i+1}^{[3]}} \Bigr) - \frac{1}{2} {\rm Li}_2 \Bigl( 1 - \frac{t_i^{[2]}
t_{i+3}^{[2]}}{t_i^{[3]}
  t_{i+2}^{[3]}} \Bigr) + \frac{3}{2} \zeta_2 \biggr],
\end{align}
where $t_i^{[r]} = (p_i + ... + p_{i+r-1})^2$ are the Mandelstam kinematical invariants.

\subsection{Light-like Wilson loops}\label{llwl}

Let us now turn to the description of the light-like Wilson loops, which are the counterparts of
the MHV planar amplitudes in the duality relation discussed in this paper. In the $\mathcal{N}=4$
SYM theory with an $SU(N)$ gauge group they are defined as
\begin{equation}\label{W}
    W\lr{C_n} = \frac1{N}\vev{0|\,{\rm Tr}\, \textrm{P} \exp\lr{i\oint_{C_n} dx^\mu A_\mu(x)}
    |0}\,,
\end{equation}
where the gauge field $A_\mu(x)$ is integrated along the contour $C_n=\bigcup_{i=1}^n \ell_i$ made
out of $n$ light-like segments joining the cusp points $x_i^\mu$ (with $i=1,2,\ldots,n$)
\begin{equation}\label{4'}
 \ell_i=\{x^\mu(\tau_i)= \tau_i x^\mu_i +
(1-\tau_i)x_{i+1}^\mu|\, \tau_i \in[0,1] \}\,,
\end{equation}
such that the tangent vectors $\partial_{\tau_i} x^\mu(\tau_i) = x_{i,i+1}^\mu$ are light-like,
$x_{i,i+1}^2=0$. The symbol $\textrm{P}$ indicates the ordering of the $SU(N)$ indices along the
integration contour $C_n$.

The Wilson loop \re{W} is a gauge invariant quantity depending on the integration contour $C_n$. As
a function of the cusp points, it is invariant under their cyclic permutations and flips,
\begin{equation}
W(x_1,x_2,\ldots,x_n) = W(x_n,x_1,\ldots,x_{n-1}) = W(x_n,x_{n-1},\ldots,x_1)\,.
\end{equation}
If the contour $C_n$ did not have cusps, the Wilson loop would be a finite quantity in
$D=4$~\cite{P80,Brandt:1981kf,KK92}. Characteristic feature of the light-like contour $C_n$ is that
it maps to a similar contour under $SO(2,4)$ conformal transformations. As a result, a finite
Wilson loop would be a conformal invariant function of $x_i$~\cite{am07,Drummond:2007cf}. We would
like to stress that, contrary to the gluon scattering amplitudes, the Wilson loop \re{W} is defined
in configuration space and the conformal transformations act on the four-vectors $x_i^\mu$ defining
the positions of the cusp points in Minkowski space-time. Due to the presence of cusps on the
integration contour ${C}_n$, the Wilson loop \re{W} has specific ultraviolet divergences
~\cite{P80,Brandt:1981kf,KR87,KR86,KK92,KM93} which make the conformal symmetry of $W(C_n)$
anomalous.

To regularize the cusp singularities, we use dimensional reduction with $D=4-2\epsilon_{\rm
\scriptscriptstyle UV}$ and $\epsilon_{\rm \scriptscriptstyle UV}>0$ (notice the sign difference
compared to $\epsilon_{\rm \scriptscriptstyle IR}$). Like the scattering amplitude, the Wilson loop
can be  split into a divergent and a finite part,
\begin{equation}\label{W=Z+F}
\ln W(C_n) = Z_n^{\rm (WL)} + F_n^{\rm (WL)} + O(\epsilon_{\rm \scriptscriptstyle UV})\,.
\end{equation}
The divergent part $Z_n^{\rm (WL)}$ has the special form~\cite{KK92}
\begin{equation}\label{5}
Z_n^{\rm (WL)} = -\frac{1}{4}  \sum_{l\ge 1} a^l \left({\frac{\Gamma_{\rm
cusp}^{(l)}}{(l\epsilon_{\rm \scriptscriptstyle UV})^2}+ \frac{\Gamma^{(l)}}{l\epsilon_{\rm
\scriptscriptstyle UV}}}\right) \sum_{i=1}^n\lr{-x_{i,i+2}^2\mu_{\rm \scriptscriptstyle UV}^2}^{l\epsilon_{\rm \scriptscriptstyle UV}}\,,
\end{equation}
where $\Gamma_{\rm cusp}^{(l)}$ are the expansion coefficients of the cusp anomalous dimension
\re{cusp} and $\Gamma^{(l)}$  is the counterpart of the collinear anomalous dimension entering in
\re{Zn},
\begin{align}\label{W-col}
\Gamma(a) =\sum_{l\ge 1} a^l\, \Gamma^{(l)} = - 7\zeta_3 a^2 + O(a^3)\,.
\end{align}
The UV cut-off $\mu_{\rm
\scriptscriptstyle UV}^2$ is related to the dimensional regularization scale
$\mu^2$ as  (cf. \re{ge})
\begin{equation}\label{mu-UV}
\mu_{\rm \scriptscriptstyle UV}^2=\pi{\rm e}^{\gamma_E}\mu^2\,.
\end{equation}

The finite part of the Wilson loop, $F_n^{\rm (WL)}$, does not depend on the renormalization scale
$\mu_{\rm \scriptscriptstyle UV}^2$ and it is a dimensionless function of the distances $x_{ij}^2$
with $i,j=1,\ldots, n$. Since the edges of $C_n$ are light-like, $x_{i,i+1}^2=0$, the non-vanishing
distances are $x_{ij}^2$  with $|i-j|\ge 2$. Despite the fact that the conformal invariance of
$W(C_n)$ is broken by the cusp singularities, it imposes severe constraints on $F_n^{\rm (WL)}$. We
showed in \cite{Drummond:2007cf,Drummond:2007au} that  $F_n^{\rm (WL)}$ has to satisfy the
following {\it anomalous} conformal Ward identity:
\begin{equation}\label{SCWI}
K^\mu {F}_{n}^{\rm (WL)} \equiv \sum^n_{i=1} \left(2x_i^\nu x_i\cdot\partial_i - x_i^2
\partial_i^\nu\right)
 {F}_{n}^{\rm (WL)}=
 \frac12\Gamma_{\rm cusp}(a) \sum_{i=1}^n x^\nu_{i,i+1}\, \ln\lr{
 \frac{x_{i,i+2}^2}{x_{i-1,i+1}^2}} \,.
\end{equation}
The differential operator on the left-hand side is the conformal boost (special conformal
transformation) generator $K^\mu$. The right-hand side expresses the conformal anomaly due to the
cusp singularities. As explained in \cite{Drummond:2007au}, it has a universal functional form,
with the coupling dependence coming only through the cusp anomalous dimension $\Gamma_{\rm
cusp}(a)$. For $n=4$ and $n=5$ this relation is powerful enough to determine the all-loop
expressions for ${F}_{4}^{\rm (WL)}$ and ${F}_{5}^{\rm (WL)}$ up to an additive coupling-dependent
constant. The reason for this is that any potential solution of the homogeneous differential
equation $K^\mu {F}_{n}^{\rm (WL)} =0$ would be a conformal invariant. It is well known that such
invariants take the form of cross-ratios ${x^2_{ij}x^2_{kl}}/\lr{x^2_{ik}x^2_{jl}}$. It is then
immediately clear that one cannot build invariants from four or five points $x_i^\mu$ with
light-like separations $x_{i,i+1}^2=0$. However, they can be constructed starting from six points.
In particular, for the hexagon Wilson loop $W(C_6)$ there are three such cross-ratios,
\begin{equation}\label{u1u2u3}
u_1 = \frac{x_{13}^2 x_{46}^2}{x_{14}^2 x_{36}^2}, \qqqquad u_2 = \frac{x_{24}^2 x_{15}^2}{x_{25}^2
x_{14}^2}, \qqqquad u_3 = \frac{x_{35}^2 x_{26}^2}{x_{36}^2 x_{25}^2}\ .
\end{equation}
As a result, for $n\ge 6$ the general solution of the conformal Ward identity \re{SCWI} will
contain an arbitrary function of the conformal cross-ratios. This function was calculated at two
loops for $n=6$ in \cite{Drummond:2007bm}. In the present paper we give more details of this
calculation and compare the result with the corresponding six-gluon MHV amplitude.

\subsection{Duality relation}\label{dr}

In this subsection we formulate and discuss the main point of the present paper -- the proposed
duality relation between the MHV planar amplitudes $\mathcal{M}_n^{\rm (MHV)}$ and the light-like
Wilson loops $W(C_n)$.

{The conjectured duality states that in the planar $\mathcal{N}=4$ SYM theory the finite parts of the logarithms of the
gluon amplitude and of the Wilson loop are equal (up to an inessential additive constant):
\begin{equation} \label{duality-3}
{F}_n^{(\mathrm{MHV})} =  F^{(\textrm{WL})}_n + \mathrm{const} \,,
\end{equation}
upon the formal identification of the external on-shell gluon momenta in the amplitude with the
light-like segments forming the closed polygon $C_n$ (the contour of the Wilson loop),
\begin{equation} \label{duality-1}
p_i^{\mu} := x_{i}^{\mu}-x_{i+1}^{\mu}\,.
\end{equation}
Thus, the Mandelstam variables for the scattering amplitudes $t^{[j]}_i=(p_i+\ldots+p_{i+j-1})^2$ are related to the distances $x_{ij}^2$ between two cusp points on the integration contour of $W(C_n)$
as follows,
\begin{equation}\label{idmomco}
  t^{[j]}_i/t^{[l]}_k :=
{x_{i,i+j}^2}/{x_{k,k+l}^2}\,.
\end{equation}

The divergent parts of the scattering amplitudes and the light-like Wilson loops are also related to each
other but the relationship is more subtle since the two objects are defined in two different
schemes (infrared regularization for the amplitudes and ultraviolet regularization for the Wilson loops),
both based on dimensional regularization.
{}From the discussion in Subsections \ref{pmhva} and \ref{llwl} we know that the leading IR divergence of the amplitude (the coefficient of the double pole $\epsilon^{-2}_{\mathrm{IR}}$ in Eq.~\re{Zn}) coincides with the leading UV divergence of the Wilson loop (the coefficient of the double pole  $\epsilon^{-2}_{\mathrm{UV}}$ in Eq.~\re{5}), since both are controlled by the universal cusp anomalous dimension $\Gamma_{\rm cusp}(a)$.
One can also achieve the matching of the coefficients of the subleading simple poles corresponding to the (non-universal) collinear anomalous dimensions $G(a)$ and $\Gamma(a)$ from Eqs.~\re{cusp} and \re{W-col}, respectively.\footnote{We thank Paul Heslop for turning our attention to the incomplete discussion of this point in the first version of the present paper. We are also grateful to Lance Dixon for a discussion of the different physical interpretations of the IR and UV simple poles \cite{Dixon:2008gr}.  } To this end one relates the parameters
of the two different renormalization schemes as follows,
\begin{equation} \label{duality-2}
 x_{i,i+2}^2\,\mu^2_{\mathrm{UV}} :=
 t_i^{[2]}/\mu_{\mathrm{IR}}^2\e^{\gamma(a)}\,,\qqqquad
 \epsilon_{\rm \scriptscriptstyle UV} := -\epsilon_{\mathrm{IR}} \e^{\epsilon_{\mathrm{IR}}\delta(a)}\,.
%\qquad  t^{[j]}_i/t^{[l]}_k :={x_{i,i+j}^2}/{x_{k,k+l}^2}\,.
\end{equation}
Here the functions $\gamma(a)$ and $\delta(a)$ are chosen in a way to compensate the mismatch between  $G(a)$ and $\Gamma(a)$, without creating extra $\mu$-dependent finite terms. It is easy to check that these functions are the solutions to the equations
\begin{eqnarray}  \nonumber
 \gamma(a) \Gamma_{\rm cusp} (a) +\delta(a) \tilde{\Gamma}_{\rm cusp}(a) + G(a)+\Gamma(a)  &=& 0\,,
\\[3mm]
\gamma(a) \tilde \Gamma_{\rm cusp}(a) +2 \delta(a)\tilde{\tilde \Gamma}_{\rm cusp}(a) + \tilde G(a) + \tilde \Gamma(a) &=& 0\,,
\label{examatch}
\end{eqnarray}
where $\tilde \Gamma_{\rm cusp}(a) = \int_0^a\frac{da'}{a'}  \Gamma_{\rm cusp}(a')$ and $\tilde{\tilde \Gamma}_{\rm cusp}(a) = \int_0^a\frac{da'}{a'} \tilde \Gamma_{\rm cusp}(a')$, and similarly for $\tilde G(a)$ and $\tilde \Gamma(a)$.
It should be stressed that this procedure is not analogous to comparing two different renormalization schemes for the computation of the same divergent object. It is rather a change of variables (regularization parameters) which allows us to compare two different objects computed in two different schemes. Another such change of variables is the identification (\ref{duality-1}) of the particle momenta with the light-like segments on the Wilson loop contour.

The duality relation (\ref{duality-3}) was inspired by the prescription of Alday and Maldacena {for
computing} gluon scattering amplitudes at strong coupling \cite{Alday:2007he}, which essentially
recasts the amplitudes into light-like Wilson loops in the dual variables (\ref{duality-1}). {\sl A
priori}, one would expect the strong coupling relation between gluon amplitudes and Wilson loops to
receive $1/\sqrt{\lambda}$ corrections, which might spoil the relation at weak coupling
\cite{Alday:2007he}. Nevertheless, in \cite{DKS07} three of us found that at one loop and for four
points the Wilson loop and the gluon amplitude agree, which lead to the idea that the duality might
also be true perturbatively. In \cite{BHT07} the duality at one loop was shown to apply also to
$n$-point amplitudes. Further evidence in favor of the duality was accumulated by the present
authors by two-loop calculations of the four-point \cite{Drummond:2007cf} and five-point
\cite{Drummond:2007au} Wilson loops, which were shown to match the corresponding gluon amplitudes
according to Eqs.~(\ref{duality-3}) and (\ref{duality-1}).}

What can we say about the proposed duality to all orders in the coupling $a$? {Supposing the
duality holds}, one immediate consequence is that $F_n^{\rm (MHV)}$, after making the change of
variables (\ref{duality-1}), should satisfy the conformal Ward identity (\ref{SCWI}). We stress the
fact that, while the conformal properties of the Wilson loop are manifest, there is no obvious
reason to expect that the matching gluon amplitudes should have any reasonable behavior under
conformal transformations acting on the particle {\it momenta}. Yet, the BDS ansatz for the MHV
amplitudes does have this unexpected conformal property. For instance, when rewritten in terms of
the dual variables $x_i$ (\ref{duality-1}), the BDS ansatz  for four \re{f4} and five \re{f5}
gluons reads as follows:
\begin{align}\label{n45}
F_4^{\rm (BDS)} &= \frac{1}{4} \Gamma_{\rm cusp}(a) \ln^2 \Bigl(
\frac{x_{13}^2}{x_{24}^2} \Bigr)+ \text{ const },\\
F_5^{\rm (BDS)} &= -\frac{1}{8} \Gamma_{\rm cusp}(a) \sum_{i=1}^5 \ln \Bigl(
\frac{x_{i,i+2}^2}{x_{i,i+3}^2} \Bigr) \ln \Bigl(
\frac{x_{i+1,i+3}^2}{x_{i+2,i+4}^2} \Bigr) + \text{ const }. \label{n45'}
\end{align}
It is easy to check that these formulae are indeed solutions of the Ward identity \re{SCWI}. In
fact, they are the unique solutions (up to an additive constant) for $n=4,5$, so they give the
all-order form for $F_4^{\rm (WL)}$ and $F_5^{\rm (WL)}$. In particular, this confirms the duality
up to three loops for $n=4$, since on the one hand, the BDS ansatz is known to be correct up to
three loops in this case, and on the other hand, the $n=4$ Wilson loop does satisfy the Ward
identity to all orders.

A hint at a possible source of this surprising conformal symmetry of the MHV amplitudes comes from
the observation that all momentum integrals entering in the expression for ${\cal M}_4$ up to four
\cite{bds05,Bern:2006ew} (and possibly even five \cite{5loops}) loops are of the `pseudo-conformal'
(also referred to as `dual conformal') type \cite{Drummond:2006rz,DKS07}.\footnote{The conformal
properties of ladder (scalar box) multiloop integrals were first described by Broadhurst (see the
first reference in \cite{Drummond:2006rz}).} The conformal properties of such momentum integrals
are revealed by rewriting them in terms of `dual coordinates' according to \re{duality-1}. If their
external legs are taken off shell, one can remove the dimensional regulator and the integrals
become manifestly covariant under the action of the conformal group $SO(2, 4)$ on the dual
coordinates (i.e., on the particle momenta). It should be made very clear that this symmetry of the
integrals (broken on shell by the IR divergencies) is not related to the original conformal
symmetry of the ${\cal N}=4$ SYM theory, acting on the gluon fields in the configuration space. It
is also important to realize that this unexpected property of the momentum integrals contributing
to the MHV gluon amplitudes does not automatically imply that the amplitude should satisfy the
anomalous (dual) conformal Ward identity \re{SCWI}. This is only true for some very special
combinations of such pseudo-conformal integrals (for example, all the integrals in the four-gluon
amplitude up to four (five) loops appear with coefficients $\pm 1$
\cite{Bern:2006ew,5loops,DKS07,Cachazo:2008dx}). The exact link between the pseudo-conformal
property of the integrals and the dual conformal behavior of the amplitude is still unclear.

At this stage one might suspect that the observed duality relation for $n=4,5$ is true only
because both objects have the same conformal symmetry. To put it differently, the proposed duality
might be reduced to the weaker (but still highly non-trivial) statement that the MHV gluon
amplitudes have dual conformal symmetry. The first real test of the stronger form of the duality
(\ref{duality-3}) is provided by the case $n=6$.

The BDS ansatz for the six-gluon amplitude \re{f6}, rewritten  in the dual
variables $x_i$, reads
\begin{align}\label{BDS6pointxnotation}
F_6^{\rm (BDS)}  =  \frac{1}{4} \Gamma_{\rm cusp}(a) \sum_{i=1}^{6} &\bigg[
    - \ln\Bigl(
\frac{x_{i,i+2}^2}{x_{i,i+3}^2} \Bigr)\ln\Bigl(
\frac{x_{i+1,i+3}^2}{x_{i,i+3}^2} \Bigr)
\nonumber\\
 &  +\frac{1}{4} \ln^2 \Bigl( \frac{x_{i,i+3}^2}{x_{i+1,i+4}^2}
\Bigr)   -\frac{1}{2} {\rm{Li}}_{2}\Bigl(1-
 \frac{x_{i,i+2}^2  x_{i+3,i+5}^2}{x_{i,i+3}^2 x_{i+2,i+5}^2}
\Bigr) \bigg] + \rm{const}\,.
\end{align}
Again, we can immediately verify that this is a solution of the Ward identity (\ref{SCWI}).
However, unlike the cases $n=4,5$ where the solutions (\ref{n45}), (\ref{n45'}) are unique up to an
additive constant, this is no longer true for $n\geq 6$. As explained at the end of
Subsection~\ref{llwl}, for $n\geq6$ the general solution of  (\ref{SCWI}) can contain an arbitrary
function of the conformal cross-ratios (and of the coupling constant). Thus, for $n=6$, we can
expect a possible deviation from the BDS ansatz of the form
\begin{equation} \label{duality-6}
F_6^{\rm (WL)}=F_6^{\rm (BDS)}+R_6(u_1,u_2,u_3;a)\,.
\end{equation}

The one-loop Wilson loop calculation of \cite{BHT07} has shown that the `remainder' function $R_6$
is just a constant at one loop, which is a confirmation of the Wilson loop/scattering amplitude
duality going beyond the scope of conformal symmetry. However, one might suspect just a low
loop-order `accident'. The point is that the function $R_6$ must satisfy a further, rather powerful
constraint, the so-called collinear limit \cite{Bern:1994zx,Anastasiou:2003kj} (see also
Subsection~\ref{hwlvbds} for further comments). It could be that due to the limited choice of loop
integrals at this low perturbative level, the function \re{BDS6pointxnotation} made of them is the
only one satisfying both the conformal Ward identity \re{SCWI} and the collinear limit. If so, at
some higher perturbative level new functions with these properties might appear which could spoil
the BDS ansatz and/or the Wilson loop/scattering amplitude duality.

Indeed, our recent explicit $n=6$ Wilson loop calculation \cite{Drummond:2007bm} has shown that at
two loops there exists a non-trivial `remainder' function $R_6$ satisfying both conditions. The
crucial test then was to compare the results of our Wilson loop calculation with a  parallel
two-loop six-gluon amplitude calculation, in order to check whether the proposed duality between
Wilson loops and gluon amplitudes continues to hold at this level (which, if true, automatically
implies the breakdown of the BDS ansatz). {The results of the six-gluon calculation have become
available very recently \cite{parallel}. As we demonstrate in Section~\ref{drfsgma}, the detailed
numerical comparisons of the two calculations shows that indeed the BDS ansatz fails at two loops
whereas the {duality with Wilson loops is preserved},
\begin{equation} \label{duality-7}
F_6^{\rm (MHV)}=F_6^{\rm (WL)}-c_6(a)\,.
\end{equation}
We consider this very strong evidence that the duality should hold to all orders in the coupling,
although the `remainder' function $R_n$ is likely to receive corrections at each loop order.

\section{Light-like hexagon Wilson loop}

The two-loop calculation of the light-like hexagon Wilson loop $W(C_6)$ goes along the same lines
as the analysis of the rectangular ($n=4$) and pentagonal ($n=5$) Wilson loops $W(C_n)$ performed
in Refs.~\cite{Drummond:2007cf,KK92}, where the interested reader can find the details of the
technique employed.

As was explained in detail in \cite{Drummond:2007cf}, the two-loop calculation of $W(C_6)$ can be
significantly simplified by making use of the non-Abelian exponentiation property of Wilson
loops~\cite{Gatheral}. In application to $W(C_6)$, it can be formulated as follows:
\begin{equation}\label{W-decomposition}
\ln W(C_6) = \frac{g^2}{4\pi^2}C_F\,
w^{(1)}  +  \lr{\frac{g^2}{4\pi^2}}^2 C_F N\, w^{(2)}  + O(g^6)\,,
\end{equation}
where $C_F=(N^2-1)/(2N)$ is the quadratic Casimir of the $SU(N)$ in the fundamental representation
and $w^{(1,2)}$ are dimensionless ($N-$independent) functions of the distances $x_{ij}^2$ and UV
cut-off $\mu^2_{\rm \scriptscriptstyle UV}$. In calculating $w^{(2)}$ we do not rely on the planar
limit. As a result, equation \re{W-decomposition} is exact in $N$. \footnote{Up to three loops, the
color factors in front of $w^{(n)}$ on the right-hand side of \re{W-decomposition} have the form
$C_F N^n$ and the only source of non-planar corrections is  the subleading term in the expression
for the Casimir  $C_F=N/2 -1/(2N)$. Starting from four loops, the color factors are not expressible
in terms of simple Casimir operators and receive genuine nonplanar corrections~\cite{Gatheral}.} We
will take the planar limit later on, when we need to compare our result to the planar MHV
amplitude. The one-loop correction $w^{(1)}$ was found in \cite{BHT07}. Its divergent part
coincides with the one-loop contribution to $Z_n$, Eq.~\re{Zn}, and its finite part coincides up to
an additive constant  with the BDS ansatz \re{BDS} and {satisfies} the duality relation
\re{finiteduality} at one loop.

For finite $N$ the two-loop corrections to $W(C_6)$ involve two different color factors $C_F^2$ and
$C_F N$. According to \re{W-decomposition}, the coefficient in front of $g^4C_F^2/(4\pi^2)^2$ in
the two-loop expression for  $W(C_6)$ is given by $\lr{w^{(1)}}^2/2$. Therefore, in order to
determine the function $w^{(2)}$ one has to calculate the contribution to $W(C_6)$ only from
two-loop diagrams containing `maximally non-Abelian' color factor $C_F N$. This property
significantly  reduces the number of relevant two-loop diagrams. The function $w^{(2)}$ is gauge
invariant, so in order to simplify the calculation we shall employ the Feynman gauge. In this
gauge, some of the `maximally non-Abelian' diagrams like those where  both ends of a gluon are
attached to the same light-like segment vanish by virtue of $x_{j,j+1}^2 =0$. The corresponding
Feynman diagrams have the same topology as for the rectangular $(n=4)$ and pentagon $(n=5)$ Wilson
loops $W(C_n)$ and they can easily be identified by applying the selection rules formulated in
Ref.~\cite{Drummond:2007cf}.

\begin{figure}
\psfrag{1}[cc][cc]{\small (a)}\psfrag{2}[cc][cc]{
(b)}\psfrag{3}[cc][cc]{(c)}\psfrag{4}[cc][cc]{(d)}
\psfrag{5}[cc][cc]{(e)}\psfrag{6}[cc][cc]{(f)}\psfrag{7}[cc][cc]{(g)}\psfrag{8}[cc][cc]{(h)}
\psfrag{9}[cc][cc]{(i)}\psfrag{10}[cc][cc]{(j)}\psfrag{11}[cc][cc]{(k)}
\psfrag{12}[cc][cc]{(r)}\psfrag{13}[cc][cc]{(m)}\psfrag{14}[cc][cc]{(n)}
\psfrag{15}[cc][cc]{(o)}\psfrag{16}[cc][cc]{(p)}\psfrag{17}[cc][cc]{(q)} \psfrag{18}[cc][cc]{(l)}
\psfrag{19}[cc][cc]{(s)}\psfrag{20}[cc][cc]{(t)}\psfrag{21}[cc][cc]{(u)} \psfrag{x1}[rc][cc]{$x_1$}
\psfrag{x2}[rc][cc]{$x_6$} \psfrag{x3}[rc][cc]{$x_5$} \psfrag{x4}[lc][cc]{$x_4$}
\psfrag{x5}[lc][cc]{$x_3$} \psfrag{x6}[lc][cc]{$x_2$}
\centerline{{\epsfxsize14cm \epsfbox{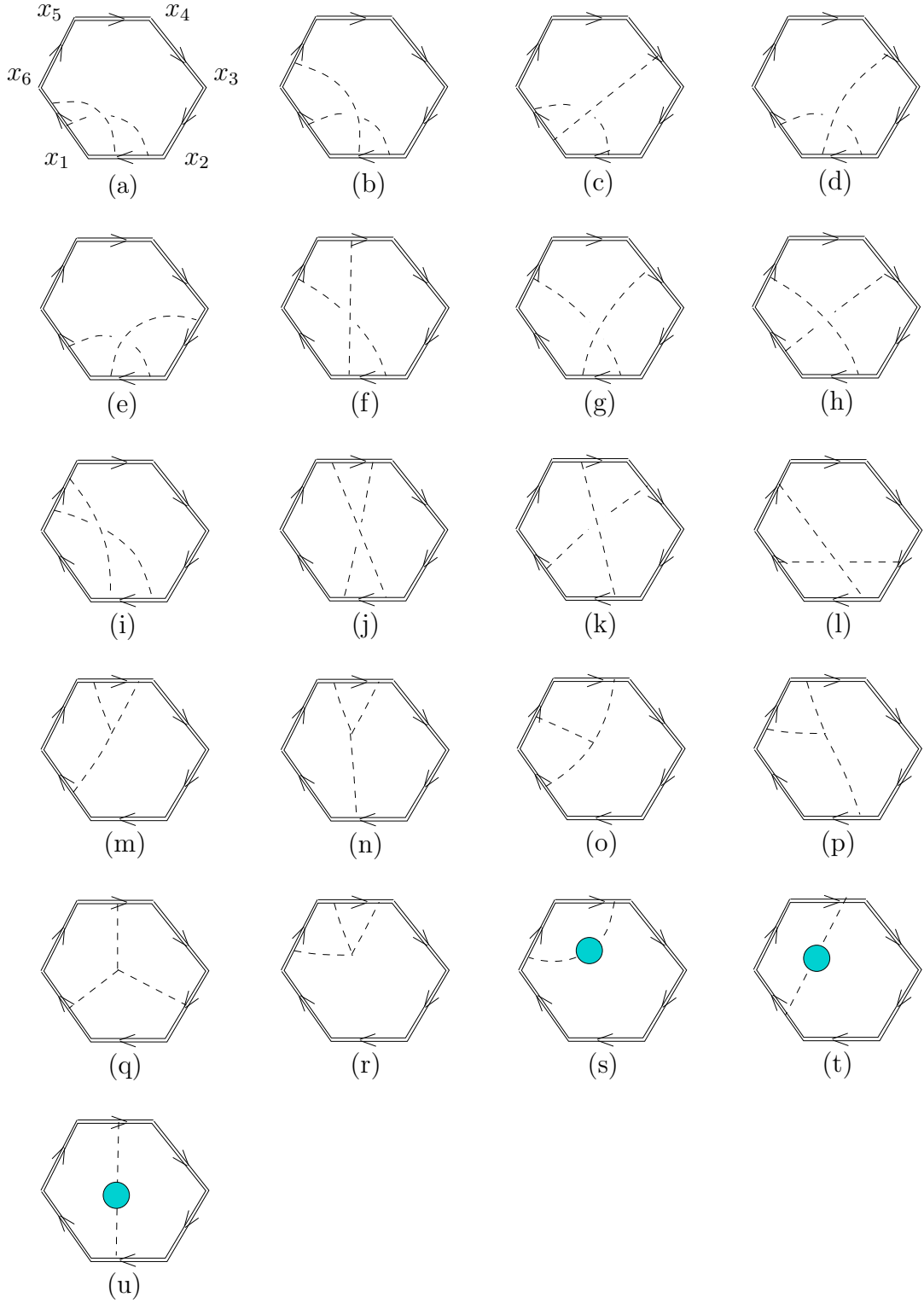}}} \caption[]{\small
  The maximally non-Abelian
Feynman diagrams of different topology contributing to $F_6^{\rm
  (WL)}$. The double lines depict
the integration contour $C_6$, the dashed lines the gluon propagator
and the blob the one-loop
polarization operator. } \label{all-diags}
\end{figure}

To summarize,  in Fig.~\ref{all-diags} we list all non-vanishing two-loop diagrams of different
topologies contributing to $w^{(2)}$.  The diagrams shown in Figs.~\ref{all-diags}(m) -- (r)
involve the three-gluon interaction vertex of the $\mathcal{N}=4$ SYM Lagrangian. Their color
factors equal $C_F N$, and therefore they contribute to the function $w^{(2)}$ in
\re{W-decomposition}. The non-planar diagrams in Figs.~\ref{all-diags}(a) -- (l) involve two free
gluon propagators and their color factors equal $C_F(C_F-N/2)$. To identify their contribution to
$w^{(2)}$, we have to retain the maximally non-Abelian part only, that is, to replace their color
factors by $C_F(C_F-N/2)\to -C_F N/2$. Finally, the diagrams in Figs.~\ref{all-diags}(s) --(u)
involve the one-loop correction to the gluon propagator with the blob denoting gauge
fields/gauginos/scalars/ghosts propagating along the loop. Their color factors equal $C_F N$ and
they directly contribute to $w^{(2)}$. To preserve supersymmetry, we evaluate these diagrams within
the dimensional reduction scheme. The two-loop correction $w^{(2)}$ is given by the sum over the
individual diagrams shown in Fig.~\ref{all-diags} plus crossing symmetric diagrams obtained either
by cyclic permutations of the cusp points $(1,2,\ldots,n)\mapsto (n,1,\ldots,n-1)$, or by flips
$(1,2,\ldots,n)\mapsto (n,n-1,\ldots,1)$ with $n=6$.

To compute the diagrams shown in Fig.~\ref{all-diags}, we employ the technique developed in
Refs.~\cite{KR86,KK92}. The calculation goes along the same lines as for the rectangular and
pentagonal Wilson loops \cite{DKS07,Drummond:2007cf,Drummond:2007au} and the result can be
summarized as follows. It is convenient to expand the contribution of each diagram in powers of
$1/\epsilon$ and separate the UV divergent and finite parts as
\begin{equation}\label{para}
w^{(2)} =   \sum_{\alpha}
\bigg\{\frac12\left(\frac1{\epsilon^4}A_{-4}^{(\alpha)}
+ \frac1{\epsilon^3}A_{-3}^{(\alpha)} + \frac1{\epsilon^2}A_{-2}^{(\alpha)} +
\frac1{\epsilon}A_{-1}^{(\alpha)}\right)\sum_{i=1}^6
\lr{{-x_{i,i+2}^2}\,{\mu_{\rm\scriptscriptstyle UV}^2}}^{2\vep} +
A_0^{(\alpha)}\bigg\} + O(\epsilon)\,,
\end{equation}
where $\epsilon=\epsilon_{\rm\scriptscriptstyle UV}$  and the scale $\mu_{\rm\scriptscriptstyle
UV}^2$ is defined in \re{mu-UV}. {Here the index $\alpha$ runs over the two-loop Feynman diagrams}
shown in Fig.~\ref{all-diags}(a)--(u) and the factor $1/2$ has been inserted for later convenience.
For each diagram, the coefficient functions $A^{(\alpha)}_{-n}$ depend on dimensionless ratios of
distances $x_{i,i+2}^2$ and $x_{i,i+3}^2$ (with $i=1,\ldots,6$ and the periodicity condition
$i+6\equiv i$). We would like to stress that the contribution of each individual diagram to
$w^{(2)}$, or equivalently the functions $A^{(\alpha)}_{-n}$, are gauge dependent and it is only
their sum on the right-hand side of \re{para} that remains gauge invariant. We note that many of
the coefficient functions $A_{-n}^{(\alpha)}$ can be cross-checked against similar contributions to
$W({C}_5)$ and $W({C}_4)$ \cite{Drummond:2007cf,Drummond:2007au}, which can be obtained by
shrinking one and two segments, respectively, of the contour $C_6$ to a point.

\subsection{Divergent part}

Summarizing our results for the UV divergent part of $W(C_6)$, we find that the coefficients
$A_{-4}^{(\alpha)}$, $A_{-3}^{(\alpha)}$ and $A_{-2}^{(\alpha)}$ of the poles on the right-hand
side of \re{para} are given by:

\begin{itemize}

\item $O(\epsilon^{-4})$ terms only come from the two Feynman diagrams
shown in Figs.~\ref{all-diags}(a) and \ref{all-diags}(r).
\begin{equation}\label{A4}
A_{-4}^{\rm (a)}=-\frac1{16}\,,\qqqquad A_{-4}^{\rm (r)}= \frac1{16}
\end{equation}

\item  $O(\epsilon^{-3})$ terms only come from the two Feynman diagrams
shown in Figs.~\ref{all-diags}(r) and \ref{all-diags}(s).
\begin{equation}\label{A3}
A_{-3}^{\rm (r)}=-\frac1{8}\,,\qqqquad A_{-3}^{\rm (s)}=\frac1{8}
\end{equation}

\item  $O(\epsilon^{-2})$ terms only come from the Feynman diagrams
shown in Figs.~\ref{all-diags}(a), (e), (o), (r) and (s) .
\begin{equation}\label{A2}
A_{{-2}}^{\rm (a)}=-{\frac {\pi^2}{96}}\,,\qquad A_{{-2}}^{\rm (e)}=-\frac{\pi^2}{24}\,,\qquad
A_{{-2}}^{\rm (o)}= \frac{{\pi }^{2}}{48} \,,\qquad A_{{-2}}^{\rm (r)}=-\frac14+{\frac
{5}{96}}\,{\pi }^{2}\,,\qquad A_{{-2}}^{\rm (s)}=\frac14
\end{equation}
 \end{itemize}
All these contributions match the corresponding ones for $W(C_4)$ and $W(C_5)$.
\begin{itemize}
\item
$O(\epsilon^{-1})$ terms come from the Feynman diagrams shown in
Figs.~\ref{all-diags}(a)--\ref{all-diags}(e),\ref{all-diags}(m) -- \ref{all-diags}(p) and
\ref{all-diags}(r) -- \ref{all-diags}(u).
\end{itemize}
The expressions for $A_{-1}^{\rm (a)}$, $A_{-1}^{\rm (e)}$, $A_{-1}^{\rm (o)}$ and $A_{-1}^{\rm
(s)}$ are the same as for $W(C_4)$ and $W(C_5)$, while the remaining $A_{{-1}}-$coefficients  are
given by complicated functions of the distances $x_{i,i+2}^2$ and $x_{i,i+3}^2$. To determine them
we apply the `subtraction procedure' described in~\cite{Drummond:2007au}. As we will see
shortly,  these functions cancel against each other in the sum of all diagrams leading to
\begin{equation}\label{A-minus}
\sum_\alpha A_{-1}^{(\alpha)} =\frac{7}{8} \zeta_3\,.
\end{equation}
%%%%%%%%%%%%%%%%%%%%%%%%%%%%%%%%%%%%%%%%%%%%%%%%%%%%%%%%%%%%%%%%%%%%%%%%%%
Substituting the relations \re{A4} -- \re{A-minus} into \re{para} we find that the divergent part
of $w^{(2)}$ is given by
\begin{equation}\label{w2-polygon}
w^{(2)} =\left\{\epsilon^{-2}\frac{\pi^2}{96}+\epsilon^{-1}
\frac{7}{16} \zeta_3 \right\}\sum_{i=1}^6
(-x_{i,i+2}^2\,\mu^{2})^{2\epsilon}+
O(\epsilon^0)\,.
\end{equation}
As we have shown in \cite{Drummond:2007au}, a similar relation holds for an arbitrary $n-$gon
light-like Wilson loop $W(C_n)$ (with $n\ge 4$).

In the planar limit, we take into account that $C_F=N/2$ and rewrite \re{W-decomposition} as
\be
\ln W(C_6) = a w^{(1)} + 2 a^2 w^{(2)} + O(a^3)\,.
\ee
Combining it with \re{w2-polygon} we see that the divergent part of $\ln W(C_6)$ is of the expected
form (\ref{5}).

Let us now return to \re{A-minus} and specify the contribution of individual diagrams. {We
introduce, following \cite{Drummond:2007au},} the auxiliary Feynman diagrams shown in
Fig.~\ref{Fig:aux}. These diagrams involve a new fake `interaction vertex' for three gluons. Notice
that one of the gluons in Fig.~\ref{Fig:aux} is attached to a corner of the hexagon, while the
positions of the two remaining gluons are integrated over {one adjacent segment and one
non-adjacent segment}. By definition~\cite{Drummond:2007au}, the Feynman integrals associated with
the two diagrams shown in Fig.~\ref{Fig:aux} are
\begin{align}\label{I-aux}
I_{\rm aux}^{\rm (a)} &= \frac{1}2 g^4 C_F N (x_{12} \cdot x_{56}) \int_0^1 d\tau_1 \int_0^1
d\tau_5\, J(x_1,x_2+ \tau_1 x_{12},x_6+ \tau_5x_{56})\,,
\\ \notag
I_{\rm aux}^{\rm (b)} &=\frac{1}2 g^4 C_F N (x_{61} \cdot x_{34}) \int_0^1 d\tau_3 \int_0^1
d\tau_6\, J(x_1+\tau_1 x_{61},x_1, x_4+\tau_3 x_{34})\,,
\\ \notag
I_{\rm aux}^{\rm (c)} &=\frac{1}2 g^4 C_F N (x_{12} \cdot x_{34}) \int_0^1 d\tau_1 \int_0^1
d\tau_3\, J(x_1,x_2+\tau_1 x_{12}, x_4+ \tau_3 x_{34})\,,
\end{align}
where we have used the parameterization \re{4'} of the integration contour $C_6$ and have
introduced the notation for the auxiliary function~\cite{KM93,Drummond:2007au}
\begin{equation}\label{J-integral}
J(z_1,z_2,z_3) = -i  (\mu^2_{\rm \scriptscriptstyle
  UV})^{-\epsilon} \int d^{4-2\epsilon} z\, G(z-z_1) G(z-z_2)
G(z-z_3)\,,
\end{equation}
with the gluon propagator $G(x)$ defined in \re{G}.  $J(z_1,z_2,z_3)$ is a symmetric function of
the three points $z_i^\mu$ (with $i=1,2,3$) in Minkowski space-time.

\begin{figure}%[h]
\psfrag{x1}[cc][cc]{$x_1$} \psfrag{x2}[cc][cc]{$x_2$}
\psfrag{x3}[cc][cc]{$x_3$}
\psfrag{x4}[cc][cc]{$x_4$} \psfrag{x5}[cc][cc]{$x_5$}\psfrag{x6}[cc][cc]{$x_6$}
\psfrag{2}[cc][cc]{(a)} \psfrag{3}[cc][cc]{(b)}
\psfrag{4}[cc][cc]{(c)} \psfrag{d}[cc][cc]{(d)}
\psfrag{e}[cc][cc]{(e)} \psfrag{f}[cc][cc]{(f)}
\psfrag{g}[cc][cc]{(g)} \psfrag{h}[cc][cc]{(h)}
\psfrag{i}[cc][cc]{(i)} \psfrag{j}[cc][cc]{(j)}
\psfrag{k}[cc][cc]{(k)} \psfrag{l}[cc][cc]{(l)}
\psfrag{m}[cc][cc]{(m)} \psfrag{n}[cc][cc]{(n)} \psfrag{o}[cc][cc]{(o)}
\centerline{{\epsfysize4cm \epsfbox{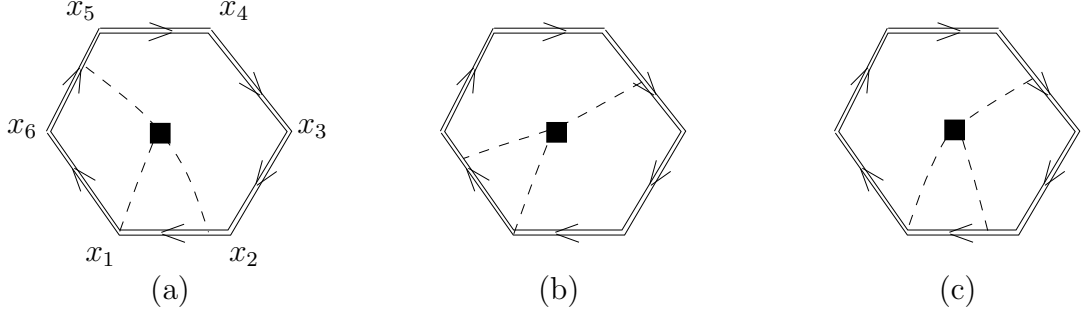}}} \caption[]{\small
  The auxiliary Feynman
diagrams defined in \re{I-aux}. The double line depicts the
  integration contour $C_6$, the dashed
line the gluon propagator and the box the fictitious three-gluon
  vertex \re{J-integral}.}
\label{Fig:aux}
\end{figure}

We observe that on the right-hand side of \re{I-aux} two of the points, say $z_2$ and $z_3$,   are
separated by a light-like interval, $z_{23}^2 \equiv (z_2-z_3)^2=0$. In this case,
$J(z_1,z_2,z_3)$ develops a simple pole in $\epsilon$ (see Eq.~\re{J-pole} for the explicit
expression). It originates from the integration in \re{J-integral} over $z^\mu$ approaching the
light-like direction defined by the vector $(z_2-z_3)^\mu$, so that the distances $(z-z_2)^2$ and
$(z-z_3)^2$ vanish simultaneously and the two propagators on the right-hand side of \re{J-integral}
become singular. As a consequence,  the integrals $I_{\rm aux}^{\rm (a,b,c)}$ have a simple pole in
$\epsilon$. Then, we add together the crossing symmetric diagrams of the same topology as in
Fig.~\ref{Fig:aux} and expand their contributions in $\epsilon$ similar to \re{para},
\begin{align}\label{M0}
I_{\rm aux}^{\rm (\alpha)} + \text{(cross-symmetry)} = \lr{\frac{g^2}{4\pi^2}}^2 C_F
N\left[\frac1{2\epsilon}M_{-1}^{\rm (\alpha)} \sum_{i=1}^6 \lr{{-x_{i,i+2}^2}\,{\mu_{\rm
      \scriptscriptstyle UV}^2}}^{2\vep} + M_{0}^{\rm (\alpha)} +
O(\epsilon)\right]\,,
\end{align}
with $\alpha=\{a,b,c\}$. Here $M_{-1}^{\rm (\alpha)}$ and $M_{0}^{\rm
  (\alpha)}$ are complicated functions of the distances $x_{jk}^2$. We
will not need their
explicit form for our purposes.

The crucial observation made in \cite{Drummond:2007au} is that the coefficient functions
$A_{-1}^{(\alpha)}$ corresponding to the various diagrams shown in Fig.~\ref{all-diags} can be
expressed in terms of the three functions $M_{-1}^{\rm (a)}$, $M_{-1}^{\rm (b)}$ and $M_{-1}^{\rm
(c)}$. In Appendix A we illustrate the underlying mechanism by presenting a detailed calculation of
the diagram shown in Fig.~\ref{all-diags}(c). Repeating the similar analysis for the remaining diagrams,
we obtained the following expressions for the coefficient functions $A_{-1}^{(\alpha)}$:
\begin{align}\label{A-1}
& A_{{-1}}^{\rm (a)}=-\frac1{24}\zeta_3\,, & &  A_{-1}^{\rm (b)} = 2
  M_{-1}^{\rm (a)}\,, & & A_{-1}^{\rm (c)} =  2 M_{-1}^{\rm (b)}\,,
\\ \nonumber
 & A_{-1}^{\rm (d)} =2 M_{-1}^{\rm (c)}\,, & & A_{{-1}}^{\rm (e)}=\frac12
\zeta_3\,,
  & & A_{-1}^{\rm (m)} +
A_{-1}^{\rm (t)} =  -M_{-1}^{\rm (a)}-M_{-1}^{\rm (c)}\,,
\\\nonumber
& A_{-1}^{\rm (n)} +A_{-1}^{\rm (u)}=-M_{-1}^{\rm (b)}\,,   & & A_{-1}^{\rm (o)} =
-M_{-1}^{\rm (a)}+\frac18 \zeta_3\,,
 & &  A_{-1}^{\rm (p)} = -M_{-1}^{\rm (b)}-M_{-1}^{\rm (c)}\,,
 \\\nonumber
 & A_{{-1}}^{\rm (r)}+A_{{-1}}^{\rm (s)}={\frac {7}{24}} \zeta_3\,.
 \end{align}
The functions $M_{-1}^{\rm (a)}$, $M_{-1}^{\rm (b)}$ and $M_{-1}^{\rm (c)}$ can be expressed in
terms of polylogarithms of degree $3$. Assigning degree 3 to the constant $\zeta_3$, we can say
that the expressions appearing in \re{A-1} have the same degree of `transcendentality'
\footnote{Here we follow the nomenclature which has become common in the literature after the
proposal of \cite{KLOV}.}. Adding them together we find that the functions $M_{-1}^{\rm (a)}$,
$M_{-1}^{\rm (b)}$ and $M_{-1}^{\rm (c)}$ cancel in the sum of all diagrams and we recover the
relation \re{A-minus}.

To summarize, we have demonstrated by an explicit two-loop calculation that the divergent part of
the hexagon Wilson loop has the expected form \re{5} with the cusp and collinear anomalous
dimensions given by \re{cusp} and \re{W-col}, respectively.

\subsection{Finite part}

The finite part of $W(C_6)$ receives contributions from all Feynman diagrams shown in
Fig.~\ref{all-diags}.  Substituting \re{para} into \re{W-decomposition} and comparing with
\re{W=Z+F}, we find the two-loop contribution to $F_{6}^{\rm (WL)} =\sum_{p\ge 1} a^p F_{6;p}^{\rm
(WL)} $,
\be\label{F62}
F_{6;2}^{\rm (WL)} = 2 \sum_{\alpha={\rm a,\ldots, u}} A_0^{(\alpha)} \,,
\ee
where the overall factor 2 comes from $(g^2/(4\pi^2))^2 C_F N = 2 a^2$ at large $N$. Based on our
analysis of the rectangular and pentagon Wilson loops~\cite{Drummond:2007cf}, we expect that the
sum on the right-hand side of this relation should have degree of transcendentality 4.

As was already explained, the contribution of the diagrams shown in Figs.~\ref{all-diags}(a), (e),
(r) and (s) to the finite part of $\ln W(C_6)$ can be obtained from the similar expressions for the
rectangular Wilson loop~\cite{Drummond:2007cf},
\begin{align}\label{A0-analytic}
& A_{{0}}^{\rm (a)}=-\frac64\cdot{\frac {7}{2880}}\,{\pi }^{4} \,,\quad
\\ \notag
& A_{{0}}^{\rm (r)}+A_{{0}}^{\rm (s)}=\frac64\cdot{\frac {119 }{2880}}\,{\pi
}^{4}\,,\quad
\\ \notag
& A_{{0}}^{\rm (e)}=\frac{\pi^2}{96}\sum_{i=1}^6
\ln^2\lr{\frac{x_{i+1,i-1}^2}{x_{i,i-2}^2}} -\frac64\cdot \frac{19}{720}\pi^4\,,
\end{align}
where the combinatorial factor $\frac64$ accounts for the difference in the  number of diagrams of
the same topology between the hexagonal and rectangular Wilson loops. Moreover, the contribution of
the diagrams shown in Fig.~\ref{all-diags}(h), (k) and (l) factorizes into a product of two
one-loop integrals and can easily be calculated.

Let us separate the remaining diagrams  into two groups according to their behavior as $\epsilon\to
0$:
\begin{itemize}
    \item The diagrams shown in Figs.~\ref{all-diags}(f), (g), (i),
    (j), (q)  are finite;
    \item {The diagrams shown in Figs.~\ref{all-diags}(b), (c), (d),
    (m), (n), (p), (t), (u)   have a
    simple pole and the diagram in Fig.~\ref{all-diags}(o) has a
    double pole.}
\end{itemize}
For the second group of diagrams we identify their finite part  by employing the  `subtraction
procedure'~\cite{Drummond:2007au}. This amounts to subtracting  from these diagrams the auxiliary
diagrams shown in Fig.~\ref{Fig:aux} with the appropriate weights defined by the coefficients in front
of the functions $M_{-1}^{\rm (a,b,c)}$ in  \re{A-1}. In this way, we compensate the simple poles
in the above mentioned diagrams and render their contribution finite (with the only exception of
the diagram in Fig.~\ref{all-diags}(o), where an additional subtraction of the  double pole defined
in \re{A2} is required). Since the auxiliary diagrams also generate a finite part $M_{0}^{\rm
(a,b,c)}$, Eq.~\re{M0}, the subtractions will modify the finite parts of the individual diagrams,
$A_{0}^{\rm (\alpha)} \to \widehat A_{0}^{\rm (\alpha)}$,
\begin{align}\label{A0-subtracted}
&  \widehat  A_{0}^{\rm (b)} =A_{0}^{\rm (b)} - 2 M_{0}^{\rm (a)}\,, & & \hspace*{-13mm} \widehat
A_{0}^{\rm (c)} = A_{0}^{\rm (c)}- 2 M_{0}^{\rm (b)}\,,   & & \widehat A_{0}^{\rm (d)} =A_{0}^{\rm
(d)}- 2 M_{0}^{\rm (c)}\,,
\\[2mm] \nonumber
 &  \widehat A_{0}^{\rm (n)+(u)} = A_{0}^{\rm (n)} +A_{0}^{\rm (u)}+M_{0}^{\rm (b)}\,,
  & & \hspace*{-13mm}  \widehat A_{0}^{\rm (o)}  =  A_{0}^{\rm (o)} +M_{0}^{\rm (a)} \,,
&&   \widehat A_{0}^{\rm (p)} =A_{0}^{\rm (p)} +M_{0}^{\rm (b)}+M_{0}^{\rm (c)}\,,
\\[2mm] \notag
&   \widehat A_{0}^{\rm (m)+(t)}  =A_{0}^{\rm (m)} +
A_{0}^{\rm (t)} +M_{0}^{\rm (a)}+M_{0}^{\rm (c)}\,. &
\end{align}
This, however, does not affect the total sum of diagrams,
\begin{equation}\sum_{\alpha}
A_{0}^{\rm (\alpha)} = \sum_{\alpha } \widehat A_{0}^{\rm
(\alpha)}\,.
\end{equation}
By construction, the subtracted diagrams are free from UV divergences and, therefore, can be
directly evaluated in $D=4$ dimensions. In this way, we find that,
\begin{equation}
\widehat A_{0}^{\rm (n)+ (u)}=\widehat A_{0}^{\rm (m)+ (t)}=0\,.
\end{equation}
In addition, we obtain a representation for the remaining functions $\widehat A_{0}^{\rm
(b,c,d,o,p)}$ and $A_{0}^{\rm (f,g,i,j,q)}$ in the form of convergent multiple integrals.   Being
combined with the explicit expressions for the remaining functions $A_{0}^{\rm (a,h,e,k,l,r,s)}$,
they determine the two-loop correction to the finite part of the hexagon Wilson loop \re{F62}.
Their explicit expressions are lengthy and we do not present them here to save space. {As an
example, we give the detailed calculation of $\widehat A_{0}^{\rm (c)}$ in the Appendix}.

It is straightforward to evaluate numerically the sum on the right-hand side of \re{F62} for a
given set of kinematical invariants,
\begin{equation}\label{distances}
%13,14,15,24,25,35,26,36,46
K=\{x_{13}^2, x_{14}^2, x_{15}^2,x_{24}^2, x_{25}^2,
x_{35}^2,x_{26}^2,x_{36}^2,x_{46}^2\}\,,
\end{equation}
defined as the distances between the vertices of the hexagon $C_6$. We should take into account,
however, that in $D=4$ dimensions there exists a relationship between the distances \re{distances},
such that only eight of them are independent. This relation reflects the fact that the five
four-dimensional vectors  $p_i^\mu = x_i^\mu - x_{i+1}^\mu$ (with $i=1,\ldots,5$) are linearly
dependent, and hence their Gram determinant vanishes
\be\label{Gram}
G[K] = \det \| (p_i \cdot p_j) \| = 0\,,\qquad (i,j=1,\ldots,5)\,.
\ee
Using $p_i=x_{i}-x_{i+1}$, the entries of the matrix can be written as linear combinations of the
distances \re{distances}.

We would like to note that the sum on the right-hand side of \re{F62} defines a function of
$x_{ij}^2$ even for configurations $K$ which do not satisfy (\ref{Gram}). This function can be
viewed as a particular continuation of the finite part of the hexagon Wilson loop off the
hypersurface defined by (\ref{Gram}).

\section{Duality relation for the six-gluon MHV amplitude}\label{drfsgma}

In the parallel publication \cite{parallel}, an impressive two-loop calculation of the six-gluon MHV
planar amplitude in $\mathcal{N}=4$ SYM theory has been performed.  To verify the duality relation
\re{duality-7} we shall compare our results for the finite part of the hexagon Wilson loop with the
numerical results from Ref.~\cite{parallel}.

\subsection{The hexagon Wilson loop versus the BDS ansatz}\label{hwlvbds}

Let us first consider the relation between the hexagon Wilson loop and the BDS ansatz,
Eq.~\re{duality-6}. To two-loop accuracy it takes the form
\begin{align}\label{WL-BDS}
 R_{\rm W}(u_1,u_2,u_3)= F_{6;2}^{\rm (WL)} - F_{6;2}^{\rm (BDS)}  \,,
\end{align}
where $R_{\rm W}$  and  $F_{6;2}^{\rm (BDS)}$ denote the two-loop {contributions} to the remainder
function \re{discrep} and to the BDS ansatz \re{BDS}, respectively. We recall that the functions
$F_{6;2}^{\rm (WL)}$ and  $F_{6;2}^{\rm (BDS)}$ satisfy the Ward identity \re{SCWI} and $R_{\rm
W}$ is a function of the three conformal cross-ratios \re{u1u2u3} only.

The simplest way to check relation \re{WL-BDS} is to evaluate the difference $F_{6;2}^{\rm (WL)} -
F_{6;2}^{\rm (BDS)} $ for two different kinematical configurations $K$ and $K'$, related to each
other by a conformal $SO(2,4)$ transformations of the coordinates $x_i^\mu$ (with $i=1,\ldots,6$).
Since $F_{6;2}^{\rm (WL)}$ and $F_{6;2}^{\rm (BDS)}$ are dimensionless functions of the distances
$x_{ij}^2$, they are automatically invariant under translations, Lorentz rotations and dilatations
of the coordinates $x^\mu_i$. The only non-trivial transformations are the special conformal
transformations (boosts), which are combinations of an inversion, a translation and another
inversion.

Let us start with the kinematical configuration $K=K(x_{ij}^2)$, Eq.~\re{distances}, and perform an
inversion of the coordinates, $x_i^\mu \to x_i^\mu/x_i^2$, to define the new configuration
\be\label{inverse}
K' = K  \left(  {x_{ij}^2}/{(x_i^2 x_j^2)}  \right).
\ee
Since the variables $u$ \re{u1u2u3} are invariant under such transformations, $u_a[K] = u_a[K']$,
the difference  $F_{6;2}^{\rm (WL)} - F_{6;2}^{\rm (BDS)} $ should also be invariant.

As an example, let us consider six light-like four-dimensional vectors $p_i^\mu$,
\begin{align}
& p_1=(1,1,0,0)\,,&&  p_2 = (-1,p,p,0)\,,&& p_3=(1,-p,p,0)\,,
\\ \notag
& p_4=(-1,-1,0,0)\,,&& p_5=(1,-p,-p,0)\,,&& p_6=(-1,p,-p,0)\,,
\end{align}
with $p=1/\sqrt{2}$  and $\sum_i p_i^\mu=0$. These vectors define the
external momenta of the gluons in the six-gluon amplitude. Applying the duality relation
$p_i=x_i-x_{i+1}$, we evaluate the corresponding distances \re{distances},
\begin{align} \notag
K^{\rm (a)}: \quad & x_{14}^2= x_{15}^2=x_{24}^2= x_{25}^2=-2\,,\quad  x_{36}^2=-2-2\sqrt{2}\,,
\\ \label{Ka}
& x_{13}^2=x_{35}^2=x_{26}^2=x_{46}^2=-2-\sqrt{2}\,,
\end{align}
and the conformal cross-ratios \re{u1u2u3},
\be\label{u-ex}
u_1=u_3=\ft12 +\ft12\sqrt{2} \,,\quad u_2=1\,.
\ee
By construction, this kinematical configuration satisfies the Gram determinant constraint
\re{Gram}. To define the conformal transformations \re{inverse}, we choose an arbitrary reference
four-vector $x_1^\mu=(x_1^0,x_1^1,x_1^2,x_1^3)$ and reconstruct the remaining $x-$vectors {
according to} $x_{i+1}^\mu = x_{i}^\mu - p_i^\mu$ {(due to translation invariance, the Wilson loop
does not depend on the choice of $x_1^\mu$)}. Then, relation \re{WL-BDS} implies that the function
$R_{\rm W}$ evaluated for the kinematical configuration \re{inverse} should be the same as for the
original configuration $K$.

As an example, we choose $x_1^\mu=(1,1,1,1)$ and apply the conformal transformation \re{inverse} to
$K^{(a)}$ defined in \re{Ka} to obtain the new kinematical configuration
\begin{align} \notag
K^{\rm (b)}: \quad & x_{14}^2= x_{15}^2=x_{24}^2= x_{25}^2=-\ft12-\ft14\,\sqrt {2}\,,\quad
x_{36}^2=-1-\ft34\,\sqrt {2}\,,
\\ \notag
&  x_{13}^2=-\ft32-\sqrt {2}\,,\quad x_{35}^2=-\ft52-\ft74\,\sqrt {2}\,,\quad
x_{26}^2=-\ft14-\ft18\,\sqrt {2}\,,
%\quad
\\ \label{Kb}
&
x_{46}^2=-\ft38-\ft14\,\sqrt {2}\,.
\end{align}
The results of our numerical tests are summarized in Table~\ref{tab:Test}. They clearly show that
$F_{6;2}^{\rm (WL)}$ and $F_{6;2}^{\rm (BDS)} $ vary under conformal transformations whereas their
difference $R_{\rm W} = F_{6;2}^{\rm (WL)} - F_{6;2}^{\rm (BDS)} $ stays invariant.

We recall that in four dimensions the kinematical invariants \re{distances} have to verify the Gram
determinant constraint \re{Gram}. This relation $G[K]=0$ is invariant under the conformal
transformations \re{inverse}, simply because the conformal boosts map six light-like vectors
$p_i^\mu$ into another set of light-like vectors. There exist, however, certain kinematical
configurations $K'$ for which  $u_i[K] = u_i[K']$ but $G[K']\neq 0$. Since the difference function
$R_{\rm W}$ only depends on the $u-$variables, its value should be insensitive to the Gram
determinant condition~\footnote{We recall that the functions entering \re{WL-BDS} can be defined
for configurations $K'$ satisfying $G[K']\neq 0$.}. For example, consider the following kinematical
configuration
\begin{align} \notag
K^{\rm (c)}: \quad & x_{14}^2= x_{15}^2=-1\,,\quad x_{24}^2= x_{25}^2=-2\,,\quad
x_{36}^2=-2-2\sqrt{2}\,,
\\ \label{Kc}
& x_{13}^2=-1-1/\sqrt{2}\,,\quad x_{35}^2=x_{26}^2=x_{46}^2=-2-\sqrt{2}\,.
\end{align}
The corresponding conformal cross-ratios \re{u1u2u3} are given by \re{u-ex}, but $G[K^{\rm
(c)}]\neq 0$. We verified numerically that $R_{\rm W}[K^{\rm (a)}]=R_{\rm W}[K^{\rm (b)}]=R_{\rm
W}[K^{\rm (c)}]$ with accuracy $< 10^{-5}$ (see Table \ref{tab:Test}). {This observation allows us
to study the function $R_{\rm W}(u_1,u_2,u_3)$ without any reference to the Gram determinant
condition.}

\renewcommand{\baselinestretch}{1.5}
\begin{table}[th]
\begin{center}
\begin{tabular}{||c||c|c|c|| }
\hline \hline Kinematical point &
 $F_{6;2}^
{\rm (WL)}$ & $F_{6;2}^ {\rm (BDS)}$ & $R_{\rm W}$
\\[1mm]
\hline \hline   $K^{\rm (a)}$ & $-5.014825  $ & $-14.294864 $ & $9.280039$
\\
\hline   $K^{\rm (b)}$& $-6.414907 $ & $ -15.694947$ & $9.280040$
\\
\hline   $K^{\rm (c)}$ & $ -5.714868$ & $ -14.994906 $ & $ 9.280038$
\\
\hline
\end{tabular}
\end{center}
\renewcommand{\baselinestretch}{1}
\caption{ Two-loop contributions to the hexagon Wilson loop, $F_{6;2}^{\rm (WL)}$, to the BDS
ansatz, $F_{6;2}^{\rm (BDS)}$,  and to their difference, $R_{\rm W}$,  evaluated for three
kinematical configurations \re{Ka}, \re{Kb} and \re{Kc} corresponding to the same values of $u_1$,
$u_2$ and $u_3$, Eq.~\re{u-ex}. } \label{tab:Test}
\end{table}
\renewcommand{\baselinestretch}{1}

The properties of the remainder function \re{WL-BDS} were studied in our previous publication
Ref.~\cite{Drummond:2007bm}. We found that $R_{\rm W}(u_1,u_2,u_3)$ is a completely symmetric
function of $u_i$.  For $u_1=0$, $u_2=u$ and $u_3=1-u$ its value is independent of $u$,
\be\label{R=const}
R_{\rm W}(0,u,1-u) = c_{\rm W}\,.
\ee
As explained in \cite{Drummond:2007bm}, this property ensures that the hexagon Wilson loop has the
same behavior as the gluon amplitude in the collinear limit. For the Wilson loop this corresponds
to flattening one of the cusp points. This lends additional support to the duality relation between
the Wilson loops and the scattering amplitudes. We found numerically that the value of $c_{\rm W}$
is given by
\be
c_{\rm W} = 12.1756 \,.
\ee
This number has a much smaller absolute accuracy ($\sim 10^{-3}$) compared to the numbers in
Table~1 because its evaluation requires taking differences of quantities that diverge in the
collinear limit.

\subsection{The hexagon Wilson loop versus the six-gluon MHV amplitude}

We now compare numerically the function $R_{\rm W}$ defined in (\ref{WL-BDS}) with the analogous
quantity defined for the MHV amplitude,
\be \label{defRA}
R_{\rm A} = F_{6;2}^{\rm (MHV)}  - F_{6;2}^{\rm (BDS)} \,.
\ee
In order to test the duality relation (\ref{duality-7}), we have to show that
for general kinematical configurations $K$ defined in equation (\ref{distances}),
\be\label{diff-diffJJ}
R_{\rm W}[K]-R_{\rm A}[K]=c_{\rm W}  \,.
\ee
Since $R_{\rm W}$ is a function of $u_{1},u_{2},u_{3}$ only, this relation would imply that so is
$R_{\rm A}$.

To get rid of the constant $c_{\rm W}$, which, as we have seen, has lower numerical precision than
the evaluations for generic kinematics, we subtract from \re{diff-diffJJ} the same relation
evaluated for some reference kinematical configuration $K^{(0)}$,
\be
R_{\rm A}[K]  - R_{\rm A}[K^{(0)}] = R_{\rm W}[K]  - R_{\rm W}[K^{(0)}] \,.
\ee
The numerical tests of this relation for different kinematical configurations are summarized in
Table~\ref{tab:A0}.
\renewcommand{\baselinestretch}{1.5}
\begin{table}[th]
\begin{center}
\begin{tabular}{||c|c||c|c|| }
\hline \hline Kinematical point & $(u_1,u_2,u_3)$ & $R_{\rm W}  - R_{\rm W} ^{(0)} $ & $R_{\rm A} -
R_{\rm A} ^{(0)}$
\\[1mm]
\hline
\hline   $K^{(1)}$ & $(1/4,1/4,1/4)$ & $< 10^{-5}$ &
$-0.018 \pm 0.023$
\\
\hline   $K^{(2)}$& $(0.547253,0.203822,0.88127)$ & $-2.75533 $ & $-2.753\pm 0.015 $
\\
\hline   $K^{(3)}$ & $(28/17,16/5,112/85)$ & $-4.74460 $ &
$ -4.7445\pm 0.0075$
\\
\hline $K^{(4)}$  & $(1/9,1/9,1/9)$ & $4.09138$
& $4.12\pm 0.10$
\\
\hline  $K^{(5)}$  & $(4/81,4/81,4/81)$ & $ 9.72553 $
& $10.00 \pm 0.50$
\\
\hline
\end{tabular}
\end{center}
\renewcommand{\baselinestretch}{1}%
\caption{Comparison of the deviation from the BDS ansatz of the Wilson loop, $R_{\rm W}$, and of
the six-gluon amplitude, $R_{\rm A}$,  evaluated for the kinematical configurations \re{Ks}. Here,
$R_{\rm W}^{(0)} =R_{\rm W}\lr{1/4,1/4,1/4} = 13.26530$ and $R_{\rm A}^{(0)}=R_{\rm
A}\lr{1/4,1/4,1/4} =1.0937\pm 0.0057$ denote the same quantities evaluated at the reference
kinematical point $K^{(0)}$. The numerical results for $R_{\rm A}$ and $R_{\rm A}^{(0)}$ are taken
from Ref.~\cite{parallel}. } \label{tab:A0}
\end{table}%
\renewcommand{\baselinestretch}{1}%
\begin{align} \notag
& K^{(0)}: && x_{i,i+2}^2 = -1\,,\qquad x_{i,i+3}^2 = -2\,;
\\[3mm] \notag
& K^{(1)}: && x_{13}^2=-0.7236200\,,\quad x_{24}^2 = -0.9213500\,,\quad x_{35}^2=-0.2723200\,,\quad
x_{46}^2=-0.3582300\,,
\\ \notag
& && x_{15}^2=-0.4235500\,,\quad x_{26}^2= -0.3218573\,,\quad x_{14}^2=-2.1486192\,,\quad
x_{25}^2=-0.7264904\,,
\\  \notag
& && x_{36}^2=-0.4825841\,;
\\[3mm] \notag
& K^{(2)}: && x_{13}^2=-0.3223100\,,\quad x_{24}^2 = -0.2323220\,,\quad x_{35}^2=-0.5238300\,,\quad
x_{46}^2=-0.8237640\,,
\\ \notag
& && x_{15}^2=-0.5323200\,,\quad x_{26}^2= -0.9237600\,,\quad x_{14}^2=-0.7322000\,,\quad
x_{25}^2=-0.8286700\,,
\\   \notag
& && x_{36}^2=-0.6626116\,;
\\[3mm]  \notag
& K^{(3)}: && x_{i,i+2}^2 = -1\,,\qquad x_{14}^2 = -1/2\,,\quad x_{25}^2=-5/8\,,\quad x_{36}^2=-17/14\,;
\\[3mm]  \notag
& K^{(4)}: && x_{i,i+2}^2 = -1\,,\qquad x_{i,i+3}^2 = -3\,;
\\[3mm] \label{Ks}
& K^{(5)}: && x_{i,i+2}^2 = -1\,,\qquad x_{i,i+3}^2 = -9/2\,.
\end{align}
Among the six configurations in equation (\ref{Ks}) only the first four verify the Gram determinant
condition \re{Gram}. Also, the configurations $K^{(0)}$ and $K^{(1)}$ are related to each other by
a conformal transformation. This explains why the first entry in the third column of
Table~\ref{tab:A0} is almost zero, and also reflects the high precision of the numerical evaluation
for the Wilson loop. Comparing the numerical values for the six-gluon amplitude and the hexagon
Wilson loop, we observe that their finite parts coincide within the error bars. Therefore, we
conclude that the duality relation \re{finiteduality} is satisfied, at least to two loops.

\section{Conclusions}

The results presented in this paper and in the parallel work \cite{parallel} show that the
conjectured Wilson loop/MHV amplitude duality holds at two loops for six cusp points/gluons,
extending previous checks in a highly non-trivial way. We believe the evidence presented here
indicates very strongly that the duality holds to all orders in the coupling. As pointed out in
\cite{Drummond:2007bm}, this necessarily implies the failure of the BDS ansatz for six gluons and
presumably beyond. Furthermore, the fact that the Wilson loop and the MHV amplitude both differ
from the BDS ansatz by the same non-trivial function $R_{\rm W}(u_1,u_2,u_3)$ suggests that there
is much more to the proposed duality than a dual conformal symmetry of the gluon amplitude. An
issue which needs to be addressed in the near future is the analytical evaluation of the remainder
function. It could give us clues about hidden symmetries of the problem and hints for generalizing
it to higher loops, as well as to strong coupling.

While at strong coupling the duality naturally emerges from the gauge/string duality, at weak
coupling we are still lacking the understanding of its origin. Since we expect that the duality
applies only to planar amplitudes, this suggests that the best way to address it will not be the
Lagrangian formulation of ${\mathcal N}=4$ SYM. Rather, it is more likely to follow from some
effective description of planar MHV amplitudes. A possible candidate
might be the twistor approach proposed in \cite{Witten:2003nn}.

Finally, let us return to the dual conformal symmetry of the gluon amplitudes. Assuming that the
duality with Wilson loops holds, this symmetry follows from the ordinary conformal symmetry of the
light-like Wilson loops in ${\mathcal N}=4$ SYM. This does not explain, however, the origin of the
duality itself. We believe that further understanding of the duality relation will come from a
deeper investigation of the symmetries of gluon amplitudes. One might speculate that dual conformal
symmetry is just part of a larger set of symmetries yet to be discovered. One possible way to
uncover such symmetries might be the loop equations~\footnote{We would like to thank A. Gorsky for
pointing out to us the formal similarity of the anomalous Ward identity (\ref{SCWI}) with the loop
equations.} \cite{Makeenko:1979pb,P80} both at weak and strong
coupling~\cite{Drukker:1999zq,Polyakov:2000jg}. Were these symmetries powerful enough to determine
the finite parts of the planar amplitudes to all loops, one would conclude that this sector of
${\mathcal N}=4$ SYM is exactly solvable.

\section*{Acknowledgements}

We would like to thank Fernando Alday, Costas Bachas, Benjamin Basso, Iosif Bena, Alexander Gorsky,
Paul Heslop, Bernard Julia, Yaron Oz and Arkady Tseytlin  for stimulating discussions. We are
particularly grateful to Zvi Bern, Lance Dixon, David Kosower, Radu Roiban, Marcus Spradlin,
Christian Vergu and Anastasia Volovich for numerous valuable discussions and for sharing their
numerical results. This research was supported in part by the French Agence Nationale de la
Recherche under grant ANR-06-BLAN-0142.

\appendix

\setcounter{section}{0} \setcounter{equation}{0}
\renewcommand{\theequation}{\Alph{section}.\arabic{equation}}

\section{Appendix: Two loop corrections to the hexagon Wilson loop}

In this Appendix we present some details of the Wilson loop calculation. It serves to illustrate
both the technique employed and to outline the general structure of two-loop corrections to the
finite part of the hexagon Wilson loop. We will examine one diagram in detail, while for the rest we will just summarize the relevant formulae.

As our detailed example, we consider the Feynman diagram shown in Fig.~\ref{all-diags}(c). It involves two
gluon propagators and originates from the expansion of the path-ordered exponential \re{W} to
fourth order in gauge fields. The end-points of the propagators are attached to three light-like
segments $[x_6,x_1]$, $[x_1,x_2]$ and $[x_3,x_4]$. It is convenient to parameterize the points
belonging to the light-like segment $x^\mu(\tau_j)\in [x_j,x_{j+1}]$ in the following way
\be
x^\mu(\tau_j) = x_{j+1}^\mu + \tau_j \, x_{j,j+1}^\mu\,,\qquad (0\le \tau_j \le 1)
\ee
with $x_{j,j+1}^\mu=x_j^\mu-x_{j+1}^\mu$ and $x_{j,j+1}^2=0$. Then the contribution of the diagram
shown in Fig.~\ref{all-diags}(c) to the Wilson loop reads
\begin{align}\label{I3-def}
I^c =  (ig)^4 N^{-1} \tr[ t^a t^b t^a t^b] & \int_0^1 \, d\tau_1 x_{12}^{\mu_1} \int_0^1 d\tau_6\,
x_{61}^{\mu_6}  \,  G_{\mu_1\mu_6}(x_{12}(1-\tau_1)+x_{61}\tau_6)
\\\notag
 \times& \int_0^1 d\tau_3 \,x_{34}^{\mu_3} \int_0^{\tau_6}  d\tau'_6\,
 x_{61}^{\nu_6}\,  G_{\mu_3
\nu_6}(x_{13}+x_{34}(1- \tau_3) + x_{61}\tau'_6)
\end{align}
where the condition $\tau'_6 \le \tau_6$ is due to the ordering of the gluons along the path and
$G_{\mu\nu}(x-y)$ is the gluon propagator. In our calculation we employ the Feynman gauge and use
dimensional regularization with $D=4-2\epsilon$ and $\epsilon>0$, so that the gluon propagator is
given by
\be\label{G}
G_{\mu\nu}(x)  = g_{\mu\nu} G(x)\,,\qquad G(x)= - \frac{\Gamma(1-\epsilon)}{4\pi^{2}} (-
x^2+i0)^{-1+\epsilon}\lr{\mu^2 \pi }^{\epsilon}\,.
\ee
We would like to stress that the contribution of each individual Feynman diagram to the Wilson loop
is gauge dependent whereas the sum of all two-loop diagrams shown in Fig.~\ref{all-diags} is gauge
invariant. As a first step, we evaluate the color factor of $I^c$
\be
{N}^{-1}\tr[ t^a t^b t^a t^b] = C_F\lr{C_F-\ft12 N},
\ee
with $C_F=t^a t^a= (N^2-1)/(2N)$ being the quadratic Casimir of $SU(N)$ in the fundamental
representation, and verify that it involves the maximally non-Abelian part $\sim C_F N$. As
explained in Sect.~3, it is only this part of $I^c$ which contributes to $\ln W(C_6)$. Then, we make
use of the identities  $x_{j,j+1}^2=0$  to simplify $[x_{12}(1-\tau_1)+x_{61}\tau_6]^2 = 2(x_{12}\,
x_{61}) (1-\tau_1)\tau_6$ and similarly for the second propagator in \re{I3-def}. The integration
over $\tau_1$, $\tau_3$ and $\tau_6$ can easily be performed leading to
\begin{align}\label{I3}
& I^c_{\rm NA} = \left(\frac{g^2}{4\pi^2}\right)^2 C_F N
\frac{\Gamma^2(1-\epsilon)}{8\epsilon^3}(\pi\mu^2)^{2\epsilon}(-x_{26}^2)^\epsilon
\\ \notag
& \times \int_0^1 dt\,(t^\epsilon-1)\frac{(x_{46}^2-x_{36}^2)-(x_{14}^2-x_{13}^2)
}{(x_{46}^2-x_{36}^2)t+(x_{14}^2-x_{13}^2)\bar t}
%\\ \notag & \times
\left[ \left(-x_{14}^2
\bar t - x_{46}^2 t \right)^\epsilon-\left(-x_{13}^2
\bar t - x_{36}^2 t \right)^\epsilon\right]
\end{align}
with $t\equiv \tau'_6$ and $\bar t=1-t$. Here the subscript indicates that we retained only the maximal
non-Abelian part of $I^c$. We observe that $I^c_{\rm NA}$ has a single pole as $\epsilon\to 0$. It
comes from the integration region $\tau_1\to 1$.

To separate the divergent and finite parts of $I^c_{\rm NA}$ we use the subtraction procedure
described in detail in Ref.~\cite{Drummond:2007au}.  We introduce the auxiliary Feynman
diagram shown in Fig.~\ref{Fig:aux}(b),
\be\label{I3-J}
I_{\rm aux}^{\rm (b)} = \frac{1}2 g^4 C_F N \lr{x_{34}\cdot x_{61}} \int_0^1 d\tau_3 \int_0^1
d\tau_6\, J(x_1+\tau_6 x_{61},x_1,x_4+\tau_3x_{34})\,,
\ee
where $J(z_1,z_2,z_3)$ stands for three propagators joined at the same point $z$ which is
integrated out\,,
\be
J(z_1,z_2,z_3) = - i \lr{\mu^{2}}^{- \varepsilon}\int d^D z\, G(z-z_1) G(z-z_2) G(z-z_3)\,.
\label{appendixJ}
\ee
The relation \re{I3-J} involves this function evaluated for  $z_1=x_1+\tau_6 x_{61}$, $z_2=x_1$ and
$z_3=x_4+\tau_3x_{34}$, so that the points $z_1$ and $z_2$ are separated by a light-like interval.
For $(z_1-z_2)^2=0$ the calculation of $J(z_1,z_2,z_3)$ yields~\cite{KM93}
\be\label{J-pole}
J(z_1,z_2,z_3) =     (\pi \mu^2)^{2\varepsilon}\frac{\Gamma(1-2\varepsilon)}{64\pi^4
\varepsilon}\int_0^1 \frac{d\tau\, (\tau\bar \tau)^{-\varepsilon}}{[-(\tau z_{13}+\bar \tau
z_{23})^2]^{1-2\varepsilon}}
\ee
with $\bar \tau=1-\tau$. We substitute this relation into \re{I3-J}, change the integration
variable via $t=\tau_6 \tau$ and expand $I_{\rm aux}^{\rm (b)}$ in powers of $\epsilon$ to find
after some algebra,
\begin{align}\label{I3-aux}
& I_{\rm aux}^{\rm (b)} = \left(\frac{g^2}{4\pi^2}\right)^2  \frac{C_F N}{16\epsilon} \int_0^1
dt\,\ln t\,\frac{(x_{46}^2-x_{36}^2)-(x_{41}^2-x_{31}^2)
}{(x_{46}^2-x_{36}^2)t+(x_{41}^2-x_{31}^2)\bar t} \ln\frac{x_{41}^2
\bar t + x_{46}^2 t}{x_{31}^2
\bar t + x_{36}^2 t }+O(\epsilon^0)\,.
\end{align}
The integral on the right-hand side of this relation can be expressed in terms of
polylogarithms of degree $3$. Symmetrizing $I_{\rm aux}^{\rm (b)}$ with respect to cyclic
permutations of indices, $i\to i+1$, and flips, $i\to 7-i$, we arrive at \re{M0}.

Let us now compare the expressions for the two diagrams, Eqs.~\re{I3} and \re{I3-aux}. We observe
that both $I_{\rm aux}^{\rm (b)}$ and $I^c_{\rm NA}$ contain a single pole $1/\epsilon$, but it
disappears in the combination $I^c_{\rm NA}-2I_{\rm aux}^{\rm (b)}$. This suggests writing $I^c_{\rm
NA}$ as
\be\label{I3-res}
I^c_{\rm NA}=2I_{\rm aux}^{\rm (b)}+I^c_{\rm fin} + O(\epsilon)
\ee
with $I^c_{\rm fin}$ finite as $\epsilon\to 0$. Combining  \re{I3} and \re{I3-aux} we
find
\begin{align}\label{I3-fin}
I^c_{\rm fin}&= -\frac18 \lr{\frac{g^2}{4\pi^2}}^2 C_F N
\lr{\frac{x_{46}^2}{x_{26}^2}-\frac{x_{36}^2}{x_{26}^2}
-\frac{x_{14}^2}{x_{26}^2}+\frac{x_{13}^2}{x_{26}^2}}
\\ \notag & \times
 \int_0^1\frac{dt\,\ln(X_1(t)/X_2(t))}{X_1(t)-X_2(t)} \left[\frac12
 \ln t \ln\lr{ {X_1(t) X_2(t)}/{t^2}}+ \int_t^1\frac{dy}{y}\ln(1-y)
 %-\ln t \ln\frac{s_{16}}{s_{36}}
 \right],
\end{align}
where
\be
X_1(t)= \frac{x_{14}^2}{x_{26}^2}(1-t)+\frac{x_{46}^2}{x_{26}^2}t \,,\qquad X_2(t)=
\frac{x_{13}^2}{x_{26}^2}(1-t)+\frac{x_{36}^2}{x_{26}^2}t\,.
\ee
Taken together, the relations \re{I3-res}, \re{I3-fin} and \re{I3-aux} define the
contribution of the diagram shown in Fig.~\ref{all-diags}(c) to $\ln W(C_6)$. Symmetrization of
$I^c_{\rm fin}$ with respect to cyclic permutations of indices and flips yields the expression for
$\widehat A_0^{\rm (c)}$ defined in \re{A0-subtracted}.

We observe that both the divergent and finite parts of $I^c_{\rm NA}$, Eq.~\re{I3-res}, are
complicated dimensionless functions of four ratios  ${x_{46}^2}/{x_{26}^2}$,
${x_{36}^2}/{x_{26}^2}$, ${x_{14}^2}/{x_{26}^2}$ and ${x_{13}^2}/{x_{26}^2}$. As such, they are not
invariant under conformal transformations of the coordinates $x_i^\mu$ (with $i=1,\ldots,6$).
Moreover, close examination of $I_{\rm aux}^{\rm (b)}$ and $I^c_{\rm fin}$ shows that the integrals
entering the relations \re{I3-aux} and \re{I3-fin}  have transcendentality $3$ and $4$,
respectively. Assigning one unit of the transcendentality to $1/\epsilon$ we find that $I^c_{\rm NA}$
has degree of transcendentality $4$.

We will now summarize the remaining non-trivial contributions to the finite part of $\ln W_6$. Each one must be included in all of its inequivalent orientations. Diagrams (a), (r) and (s) are identical to those appearing in the calculation of the rectangular Wilson loop and they contribute only a constant contribution to the finite part. Their explicit expressions can be found in \cite{KK92,Drummond:2007cf}.
The combinations (m) + (t) and (n) + (u) also do not contribute to the finite part, as described in \cite{Drummond:2007au}. For the remaining diagrams, some of the contributions (diagrams shown in Fig.2 (h),(k),(l)) factorize into products of finite one-loop integrals given by
\begin{align} \notag
J_{jk} &= \int_0^1 \frac{dt_j dt_k\, (x_{j,j+1}x_{k,k+1})}{(x_{jk}-x_{j,j+1}t_j+x_{k.k+1}t_k)^2}
\\ \label{aux-JJ}
&=-\frac12\int_0^1dt
\frac{(x_{j+1,k+1}^2-x_{j,k+1}^2)-(x_{j+1,k}^2-x_{jk}^2)}{(x_{j+1,k+1}^2-x_{j,k+1}^2)t+(x_{j+1,k}^2-x_{jk}^2)\bar
t}\ln\frac{x_{j+1,k+1}^2 t+ x_{j+1,k}^2 \bar t}{x_{j,k+1}^2 t+ x_{j,k}^2 \bar t}.
\end{align}
For the remaining expressions we will often make use of the following shorthand notation for certain ratios of the variables $x_{ij}^2$,
\begin{align}
\alpha = \frac{x_{26}^2}{x_{13}^2}, \,\,\,\,\, \beta = \frac{x_{36}^2}{x_{13}^2},
\,\,\,\,\, \gamma = \frac{x_{14}^2}{x_{13}^2}, \,\,\,\,\, \delta =
\frac{x_{46}^2}{x_{13}^2}, \,\,\,\,\, \eta = \frac{x_{15}^2}{x_{13}^2}.
\end{align}
Sometimes it is also convenient to use the notation (recall that $p_i = x_i - x_{i+1}$)
\be
s_{ij} = 2 p_i \cdot p_j = 2 x_{i,i+1} \cdot x_{j,j+1} = x_{i,j+1}^2 + x_{i+1,j}^2 - x_{ij}^2 - x_{i+1,j+1}^2.
\ee
Further, we will often use the notation $\bar y \equiv 1- y$ for some variable $y$. We will never use the bar to refer to complex conjugation.
Using this notation, the remaining non-trivial contributions are:

\subsubsection*{Diagram (b)}
\begin{align}
I^b_{\rm fin} = -\frac18 \lr{\frac{g^2}{4\pi^2}}^2 C_F N \int_0^1
\frac{dt}{t+s_{12}/s_{26}}\left[\frac12 \ln^2 X(t) \ln t +\ln X(t) \int_t^1
\frac{dy}{y}\ln\bar y \right],
\end{align}
where $X(t) = 1/(\alpha t)+1+ s_{26}/s_{16}$.

\subsubsection*{Diagram (d)}
\begin{align}
&I^d_{\rm fin}= -\frac18 \lr{\frac{g^2}{4\pi^2}}^2 C_F N
 \int_0^1\frac{dt\,\ln(X_1/X_2)}{X_1-X_2} \left[\frac12
 \ln t \ln\lr{\frac{X_1 X_2 s_{46}^2}{t^2s_{16}^2}}+ \int_t^1\frac{dy}{y}\ln\bar y
\right],
\end{align}
where $X_1=  \lr{x_{51}^2\bar
t}/\lr{x_{41}^2-x_{46}^2-x_{51}^2}$ and $X_2= \lr {x_{41}^2\bar t + x_{46}^2
t}/\lr{x_{41}^2-x_{46}^2-x_{51}^2}$.

%\begin{align*}
%X_1= \frac{x_{51}^2\bar
%t}{x_{41}^2-x_{46}^2-x_{51}^2}
%\hspace{30pt}
%X_2= \frac{x_{41}^2\bar t + x_{46}^2
%t}{x_{41}^2-x_{46}^2-x_{51}^2}
%\end{align*}

\subsubsection*{Diagram (e)}
\begin{align}
I^e_{\rm fin} &=\left(\frac{g^2}{4\pi^2}\right)^2 C_F (-\frac12 N)\left[
-\frac{\zeta_2}{8} \ln^2\lr{\frac{x_{15}^2}{x_{26}^2}}+ \frac{19}{1440}\pi^4
\right].
\end{align}

\subsubsection*{Diagram (f)}
\begin{align}
I^f_{}   = \frac{g^4}{(4\pi^2)^2} C_F (-\tfrac{1}{2}N) \frac14 &
% J^f
%\end{align}
%\\ \notag
%\begin{align}
%J^f & =
\int_0^\mu dt
\int_0^{\tfrac{\lambda}{\mu}t} dv
\frac{1}{[\lambda+1-v][\delta-\lambda-\alpha -t -1]}
\\[3mm] \notag
& \times
 \ln
\Bigl[\frac{\lambda(\lambda + 1 - v) +
    \alpha(\lambda-v)}{\alpha(\lambda-v)}\Bigr]   \ln
\Bigl[\frac{\mu(\delta-t) -t(\alpha+\lambda)}{\mu(\lambda+\alpha+1) -
    t (\alpha+\lambda)}\Bigr],
\end{align}
where $ \lambda=\beta-\alpha-1$ and $\mu=1+\delta-\gamma-\beta $.
%\begin{equation}
%\lambda=\beta-\alpha-1, \,\,\,\,\,\,\, \mu=1+\delta-\gamma-\beta.
%\end{equation}

\subsubsection*{Diagram (g)}
\begin{align}
  I^g  = \frac{g^4}{(8\pi^2)^2} C_F (-\tfrac{1}{2}N)
%J^g
%\end{align}
%\\ \notag
%\begin{align}
%& J^g  =
&  \int_0^\rho dt
\int_0^{\tfrac{\lambda}{\rho}t} dv \frac{1}{[v-\lambda-1][\delta +
    t]}
\\ \notag &  \times
    \ln \Bigl[\frac{\lambda(\lambda + 1 - v) +
    \alpha(\lambda-v)}{\alpha(\lambda - v)} \Bigr] \ln \Bigl[\frac{t
    \eta}{\rho \delta + (\rho + \eta)t} \Bigr].
\end{align}
In the expression for $I^g$ the following notation has been used:
$ \lambda = \beta - \alpha -1$ and $\rho = \gamma - \eta - \delta $.
%\begin{equation}
%\lambda = \beta - \alpha -1 \,\,\,\,\,\, \rho = \gamma - \eta - \delta.
%\end{equation}

\subsubsection*{Diagram (h)}
\begin{align}
I^h &=\frac{g^4}{(4\pi^2)^2} C_F (-\tfrac{1}{2}N)   J_{14} J_{26}\,,
\end{align}
where $J_{jk}$ is given by \re{aux-JJ}.

\subsubsection*{Diagram (i)}

\begin{align}
 I^i   = \frac{g^4}{(8\pi^2)^2} C_F (-\tfrac{1}{2}N) &
  \int_{0}^{\lambda} dv \int_{0}^{\lambda} dt
  \frac{1}{v+\bar{\lambda}} \frac{1}{t - \alpha}
\\  \notag & \times
   \ln \Bigl[\frac{t(v +
  \bar{\lambda}) + \alpha (\lambda -v)}{\alpha (\lambda - v)} \Bigr]
  \ln \Bigl[\frac{t}{v(t-\alpha) + t \bar{\lambda} + \lambda \alpha} \Bigr].
\end{align}
In the expression for $I^i$ the following notation has been used:
$\lambda = 1 + \alpha - \beta$ and $ \bar\lambda = 1 - \lambda$.
%\begin{equation}
%\lambda = 1 + \alpha - \beta, \hspace{20pt} \bar\lambda = 1 - \lambda.
%\end{equation}

\subsubsection*{Diagram (j)}
\begin{align}
I^j = & \frac{g^4}{(8\pi^2)^2} C_F (-\tfrac{1}{2}N)
 \int_0^\mu dv \int_0^\mu dt \frac{1}{[\delta -
\beta -v][1 -\beta - t]}
\\ \notag
&  \times\ln \Bigl[\frac{t(\delta -
    \beta -v) + \beta \mu + v (1-\beta)}{\beta \mu +
    v(1-\beta)}\Bigr]
\ln \Bigl[\frac{\mu (1-t) - t (\beta -
    \delta)}{v(1-\beta-t) + \beta \mu - t(\beta - \delta)}\Bigr]   .
\end{align}
In the expression for $I^j$ the following notation has been used:
$\mu = 1-\beta-\gamma+\delta$.
%\begin{equation}
%\mu = 1-\beta-\gamma+\delta.
%\end{equation}

\subsubsection*{Diagram (k)}
\begin{align}
I^k =\frac{g^4}{(4\pi^2)^2} C_F (-\tfrac{1}{2}N)   J_{14} J_{36}.
\end{align}

\subsubsection*{Diagram (l)}
\begin{align}
I^l =\frac{g^4}{(4\pi^2)^2} C_F (-\tfrac{1}{2}N)   J_{15} J_{26}.
\end{align}

\subsubsection*{Diagram (o)}
There are several contributions to the finite part coming from this diagram. We have
\be
I^o_{\rm fin} = I^o_{{\rm fin},1} +  I^o_{{\rm fin},2} +  I^o_{{\rm fin},3}.
\ee
The first is
\begin{align}
I^o_{{\rm fin},1} &= \left(\frac{g^2}{4 \pi^2}\right)^2 \frac{C_{F} N_{c}}{16}
% J^o_1 \\
%J^o_1 &=
\int_0^1 \frac{dx}{x \bar{x}} \int_0^x dt_{1}
\int_0^{\bar{x}} dt_{2} \int_0^{1} dv \Bigl[ \left[ (\bar{a} t_{1} +
a \bar{t}_{2}  )v + b t_1 t_2 \bar{v} \right]^{-1} + (a \longleftrightarrow b) \Bigr].
\end{align}
The second part is
\be I^o_{{\rm fin},2} =
\lr{\frac{g^2}{4\pi^2}}^2\frac{C_F N}{16} (K_1 + K_2 + K_3),
\ee
\begin{eqnarray}
K_{1} &=&-2 (1-a-b) \int_0^1 dt_1 dt_3 dx dv
\frac{v}{\bar{v}} \frac{a t_3 +b t_1 }{x t_1 b - \bar{x}
t_3 a} \left[ \frac{\bar{x}}{ \bar{v} b t_1 + v \bar{x}
C}-\frac{x}{\bar{v}a t_3 + v x C} \right]\nonumber \\
K_{2} &=& (1-a-b) \int_0^1 dt_1 dt_3 dx dv \frac{1}{x B -
\bar{x} A} \bigg[ \frac{B}{B \bar{v}+v \bar{x}
C}-\frac{A}{A \bar{v}+v x C} \bigg] \nonumber \\
K_{3} &=& (1-a-b) \int_0^1 dt_1 dt_3 dx dv  \bigg\{
\frac{1}{x} \left[ \frac{\bar{x} A}{x B - \bar{x} A} \frac{1}{B
\bar{v}+v \bar{x} C} + \frac{1}{B\bar{v}+v C
}\right] \nonumber \\
 &&\hspace{5cm}+ \frac{1}{\bar{x}}\left[ -\frac{x B}{x B - \bar{x} A} \frac{1}{A
\bar{v}+v x C} + \frac{1}{A\bar{v}+v C }
\right] \bigg\}
\end{eqnarray}
Here we use the notation,
\begin{eqnarray}
a=1/\beta\,,\qquad b=\alpha / \beta \,,\qquad A = a
t_3\,,\qquad B= b t_1\,,\qquad C = a t_3 \bar{t}_1 + b t_1 \bar{t}_3 + t_1 t_3\,.
\end{eqnarray}
Finally the third part is
\begin{align}
I^o_{{\rm fin},3} =\frac{1}{32}\lr{\frac{g^2}{4\pi^2}}^2C_F N \int_0^1 dt_3
\frac{s_{26}}{s_{16} +s_{26}t_3}\ln\frac{s_{16}+(s_{12}+s_{26})t_3}{s_{12}
t_3}\ln^2{t_3}  + (s_{12}\leftrightarrows s_{16}).
\end{align}

\subsubsection*{Diagram (p)}
There are several contributions to the finite part coming from this diagram. We have
\be
I^p_{\rm fin} = I^p_{{\rm fin},1} +  I^p_{{\rm fin},2} +  I^p_{{\rm fin},3} \,.
\ee
The first part is
\begin{align}
I^p_{{\rm fin},1}&=-\frac18 \lr{\frac{g^2}{4\pi^2}}^2 C_F N  \int_0^1 dv_2
dv_3\int_0^1dx\int_0^\infty \frac{{dy}\, (p_3\cdot p_6)}{(y+1)(yx \bar x z_{12}^2
+\bar x z_{23}^2 + x z_{31}^2) },
\end{align}
with
$$
z_{12}=p_2+p_3 v_2\,,\quad z_{23}=x_{31}-p_3 v_2 - p_6 \bar v_3\,,\quad
z_{31}=p_1+p_6 \bar v_3\,.
$$
Similarly the second is
\begin{align}
I^p_{{\rm fin},2}&=-\frac18 \lr{\frac{g^2}{4\pi^2}}^2 C_F N  \int_0^1 dv_1
dv_3\int_0^1dx\int_0^\infty \frac{{dy}\, (p_2\cdot p_6)}{(y+1)(yx \bar x z_{12}^2
+\bar x z_{23}^2 + x z_{31}^2) }\,,
\end{align}
with
$$
z_{12}=p_2\bar v_1 +p_3\,,\quad z_{23}=x_{41}-p_6 \bar v_3 \,,\quad z_{31}= p_1
+p_2 v_1+p_6 \bar v_3\,.
$$
The final contribution is
\be
I^p_{{\rm fin},3} = \frac18 \lr{\frac{g^2}{4\pi^2}}^2 C_F N\int_0^1 dv_1 dv_2 dv_3
\int_0^1dx \int_0^\infty \frac{dy}{1+y}\frac{x V_{31}+\bar x V_{23}}{[y x \bar x
z_{12}^2 + \bar x z_{23}^2 + x z_{31}^2]^2}\,,
\ee
where
$$
z_{12}=p_2\bar v_1 + p_3 v_2\,,\quad z_{23}=x_{31}-p_3 v_2 - p_6 \bar
v_3\,,\quad z_{31}=p_1+p_2v_1+p_6 \bar v_3=-x_{31}-p_2\bar v_1 +p_6 \bar v_3
$$
and
\begin{align*}
V_{23}&= -2(p_2\cdot p_3) (p_6\cdot x_{31}) + 4 (p_3\cdot p_6)(p_2\cdot x_{31})
-2(p_2\cdot p_3) (p_3\cdot p_6) v_2 -4(p_3 \cdot p_6) (p_2\cdot p_6) \bar v_3
\\[2mm]
V_{31}&= -2(p_2\cdot p_3) (p_1\cdot p_6) + 4(p_1\cdot p_3) (p_2\cdot p_6) + 2(p_2
\cdot p_3) (p_2\cdot p_6) v_1 + 4(p_2\cdot p_6)(p_3\cdot p_6) \bar v_3
\end{align*}

\subsubsection*{Diagram (q)}
\begin{align}\notag
I^q =  \frac{g^4 C_F N}{128 \pi^4}& \int_0^1 dt_1 dt_3 dt_5 \delta(t_1 + t_3 + t_5 -1) \int_0^1\prod_i dv_i\,
[{t_1t_3 z_{13}^2+t_3t_5 z_{35}^2+t_5t_1 z_{51}^2}]^{-2}
\\
& \times \Gamma_3\left(\partial_{z_1},\partial_{z_3},\partial_{z_5}\right)
[{t_1t_3 z_{13}^2+t_3t_5 z_{35}^2+t_5t_1 z_{51}^2}]\,.
\end{align}
Here $\Gamma_3$ is the differential operator,
$$
\Gamma_3\left(\partial_{z_1},\partial_{z_3},\partial_{z_5}\right) =- (p_1\cdot
p_3) (p_5 \cdot(\partial_{z_3}-\partial_{z_1}))- (p_3\cdot p_5) (p_1
\cdot(\partial_{z_5}-\partial_{z_3}))- (p_5\cdot p_1) (p_3
\cdot(\partial_{z_1}-\partial_{z_5})),
$$
and the $z_{jk}^2$ variables are defined by
\begin{align*}
z_{13}^2&=(p_1\bar v_1 +p_2 + p_3 v_3)^2,
\hspace{10pt}
z_{35}^2=  (p_3 \bar v_3 + p_4 + p_5 v_5)^2,
\hspace{10pt}
z_{51}^2= (p_5 \bar v_5 + p_6 + p_1 v_1)^2 .
\end{align*}
%Finally the measure $[dt]_3$ is given by
%\be
%[dt]_3 = dt_1 dt_3 dt_5 \delta(t_1+t_3+t_5 -1).
%\ee
%%%%%%%%%%%%%%%%%%%%%%%%%%%%%%%%%%%%%%%%%%%%%%%%%%%%%%%%%%%%%%%%%%%%%

\end{document}